\documentclass{article}

\usepackage{graphicx}  
\usepackage{amsmath}   
\usepackage{amssymb}   
\usepackage{color}
\usepackage[title,titletoc,toc]{appendix}

\def\eq#1{{Eq.~(\ref{#1})}}

\def\fig#1{{Fig.~\ref{#1}}}

\def\ors{{\Omega_R(t_*)}}
\def\oms{{\Omega_m(t_*)}}
\def\ohs{{H(t_*)}}

\def\cn{{CosMIn}}

\def\cc{{cosmological\ constant}}

\def\g{{\sqrt{-g}}}

\def\om{{\Omega_m}}

\def\oh{{H_0}}

\def\orr{{\Omega_R}}

  \title{Cosmological Constant from the Emergent Gravity Perspective}
  \author{T. Padmanabhan and Hamsa Padmanabhan\\
  IUCAA, Pune University Campus,\\
  Ganeshkhind, Pune- 411 007.\\
  {\small {email: paddy@iucaa.ernet.in; hamsa@iucaa.ernet.in}}
  }

  \date{} 
 
\begin{document}
  
  \maketitle

  \begin{abstract}
  
 Observations indicate that our universe is characterized by a late-time accelerating phase, possibly driven by a cosmological constant $\Lambda$, with the dimensionless parameter $\Lambda L_P^2 \simeq 10^{-122}$, where $L_P = (G \hbar /c^3)^{1/2}$ is the Planck length. In this review, we describe how the emergent gravity paradigm provides a new insight and a possible solution to the \cc\ problem. After reviewing the necessary background material, we identify the necessary and sufficient conditions for solving the \cc\ problem. We show that these conditions are  naturally satisfied in the emergent gravity paradigm in which 
(i) the field equations of gravity are invariant under the addition of a constant to the matter Lagrangian and  (ii) the cosmological constant appears as an integration constant in the solution.  The numerical value of this integration constant can be related to another  dimensionless number (called \cn) that counts the number of modes inside a Hubble volume that cross the Hubble radius during the radiation and the matter dominated epochs of the universe. The emergent gravity paradigm suggests that \cn\ has the numerical value $ 4 \pi$, which, in turn, leads to the correct, observed value of the cosmological constant. Further, the emergent gravity paradigm 
 provides an alternative perspective on cosmology and interprets the expansion of the universe itself as a quest towards holographic equipartition. We discuss the implications of this novel and alternate description of cosmology.

 \end{abstract}
  \tableofcontents 
 
 \section{Introduction}
Several recent results suggest that the field equations of gravity have the same conceptual status as the equations of, say, elasticity or fluid mechanics, making gravity an emergent phenomenon (for a review, see \cite{tpreviews}). This paradigm provides an alternative description of spacetime evolution as being driven by the difference between the surface and bulk degrees of freedom, and can be derived from a thermodynamic variational principle that extremizes the heat content of null surfaces \cite{grtd}. This approach also provides a strikingly different perspective on cosmology in general and the \cc\ in particular. The main goal of this review is to explain how one can obtain a solution to  the cosmological constant problem  in the backdrop of the emergent perspective of gravity. 

The plan of the review is as follows: In the next section, we start with a brief overview of the conventional approach to cosmology in which the Friedmann universe is described in terms of the standard parameters like $H_0,\om,\orr,...$ etc. We then motivate and describe an alternative, \textit{epoch invariant} parameterization of cosmology in which the universe is described by a set of variables (essentially three energy densities, $\rho_{\rm inf},\rho_{\rm eq}$ and $\rho_{\Lambda}$) whose numerical values are manifestly independent of the epoch at which they are measured. We identify these as the three energy densities associated with the three phases of evolution of the universe, viz. the inflationary phase ($\rho_{\rm inf}$), the radiation-matter dominated phase ($\rho_{\rm eq}$) and  the late time accelerating phase ($\rho_{\Lambda}$). Of these three,  $\rho_{\rm inf}$ and $\rho_{\rm eq}$  are expected to be determined by high-energy physics models. However, there does not exist a physical principle which determines the extremely small value of $\rho_\Lambda L_P^4 \approx 10^{-123}$ in natural units.
 
After briefly reviewing (Sec.\ref{sec:approaches}) several approaches in the literature to handle the above cosmological constant problem, we identify, in Sec.\ref{sec:nscc},  the essential theoretical ingredients  which are needed for this purpose. 
We argue that, for a viable solution to the problem, we must have the following three features in the theory of gravity: (a) The field equations of gravity must be invariant under the addition of a constant to the matter Lagrangian. (b) The cosmological constant must appear as an integration constant in the solutions. (c) To determine its value, a new physical principle is required. 

Of these three requirements, (a) and (b) are known to be naturally satisfied in the emergent gravity paradigm, as described in Sec.\ref{sec:egandcc}. We, therefore, start with the requirement (c)  in  Sec.\ref{sec:cosmin}.
Using the epoch invariant parametrization of cosmology, we construct a dimensionless number (which we call \cn), that counts the number of modes within a Hubble volume that cross the Hubble radius from the end of inflation to the beginning of the late-time accelerating phase. We show how the introduction of \cn\ and the postulate that the numerical value of \cn\ is $4 \pi$ allow us to determine the \textit{correct, observed value} of the cosmological constant for a GUTs scale inflation and the allowed range in the matter and radiation energy densities as determined from cosmological observations. In other words, \cn\ allows the determination of the numerical value of the cosmological constant from first principles in terms of other parameters that are expected to be determined from high energy physics.

In Sec.\ref{sec:egandcc}, we examine these issues in the backdrop of the emergent gravity paradigm, in which the evolution of the universe is described as a quest towards holographic equipartition, and the field equations of gravity are invariant under the additional of a constant to the matter Lagrangian with the cosmological constant appearing as an integration constant in the solutions. Together with the introduction of \cn\ and the postulate that \cn\ $= 4 \pi$, this provides a comprehensive approach that determines the value of the cosmological constant. 
 
We use the signature $(-, +,+,+)$ and units with $c = \hbar = 1$ unless otherwise specified. All numerical values of cosmological parameters are consistent with the Planck 2013 data \cite{planck2013}.

 \section{Description of the universe in terms of cosmic constants}
 \label{sec:dynevol}
 
 To study the dynamics of the universe, it is convenient to separate the evolution of a hypothetical smooth universe from the growth of structures which arise due to the amplification of the initial perturbations through gravitational instability. The background universe can be  described by a homogeneous, isotropic Friedmann model. The density perturbations are usually characterized by the initial power spectrum (possibly generated during the inflationary  phase) and specified by an amplitude and power spectrum index. Most of this review will be concerned with the evolution of the background universe, rather than the growth of structures, and hence we will begin by rapidly reviewing the standard description of the Friedmann universe.
 
 The dynamics of the Friedmann universe is essentially characterized by a single degree of freedom $a(t)$ (called the expansion factor) and the curvature of the spatial section of the universe. Observations indicate that the spatial curvature is close to zero, and hence we can describe the universe as a spatially flat FRW line element given by
 \begin{equation}
  ds^2 = -dt^2 + a^2(t)[dx^2+ x^2 (d\theta^2+\sin^2\theta d\phi^2)]
  \label{s11.1}
 \end{equation} 
 (During most part of this review, we will assume that the universe is spatially flat; our results and discussions can be easily generalized to other cases and we will make comments about this when relevant.)
 Einstein's equations then determine $a(t)$ in terms of the energy density, $\rho(t)$, and pressure, $p(t)$, of the matter which occur in the source energy-momentum tensor $T^i_k = \text{dia} [-\rho, p,p,p]$. These are given by
 \begin{equation}
  \frac{\ddot{a}}{a} = - \left(\frac{4 \pi G}{3} \right) (\rho + 3 p) ; \ \ \left(\frac{\dot{a}}{a}\right)^2 = \frac{8 \pi G}{3} \rho(t)
  \label{s11.3}
 \end{equation}
 These equations can be manipulated to give the result: $d(\rho a^3) = -p d(a^3)$. Given an equation of state $p= p(\rho)$, this equation can be integrated to give the evolution of energy density: $\rho = \rho (a)$. Using this in the second equation in \eq{s11.3}, we can determine the evolution of the universe, viz., $a=a(t)$. Similar ideas work even when the source is made of several non-interacting components like e.g, matter, radiation, dark energy etc.
 
 What is not stressed in literature is that the above, standard description of the universe gives us \textit{complete} freedom in the evolutionary history \textit{if} we do not impose any constraint on the energy density and pressure. More precisely, given \textit{any} $a(t)$, it is possible to construct a universe by choosing a suitable equation of state $p=p(\rho)$ along the following lines. We first determine $H(t) \equiv \dot a/a$ from the given $a(t)$ and then obtain the energy density and pressure of the source through the equations 
 \begin{equation}
  \rho(t)=\frac{3H^2}{8\pi G}; \quad p(t)=- \frac{1}{8\pi G}[3H^2+2\dot H]
 \end{equation}
 Eliminating $t$ between the functions $\rho(t)$ and $p(t)$, we obtain the equation of state $p=p(\rho)$ which will consistently provide the correct dynamical evolution with the previously prescribed $a(t)$. Note that the energy density $\rho(t)$ is positive by definition in the above construction while the pressure $p(t)$ could be positive or negative. Current cosmologists do not consider negative pressure a taboo, and hence the above description allows one to create \textit{any} evolutionary history by postulating a suitable equation of state for the source.
 
 Obviously, to have a reasonable, predictive theory of the universe, we need to ensure that the matter source is made of physically relevant components. Such a model has been put together over the years and is well supported by observations. These observations indicate that our universe can be characterized by three distinct phases of evolution: An early inflationary phase, driven possibly by a GUTs scale scalar field, a phase of accelerated expansion at late time, dominated by dark energy (which is assumed to be a \cc\ in this review) and a transient phase in between, dominated by radiation and matter (which includes both dark matter and baryons). 
 
A convenient parameterization of energy densities of these components in \eq{s11.3} is required to describe the  different phases of the universe 
 in a natural manner. In the next subsection, we will briefly review the standard parameterization used in cosmology. It turns out, however, that this is not best suited to study the fundamental issues we are interested in. Hence, we will motivate and introduce   in Sec.\ref{sec:epochinv} an alternate parameterization which serves our purpose.
 
 \subsection{Standard parameterization of cosmology}
 
 To describe the evolutionary history of our universe, we conventionally use a set of parameters which could be taken as the following. To begin with, 
the first and the last phases 
have approximately constant Hubble radii, $H_{\rm inf}^{-1}$ and $H_\Lambda^{-1}$, related to the respective constant rates of expansion of the universe during these phases. While the Hubble radius $H_{\rm inf}^{-1}$ of the inflationary phase is related to the mechanism driving inflation, the Hubble radius during the late-time accelerating phase can be related to a cosmological constant $\Lambda$ with $\Lambda \equiv  3 H_\Lambda^2$. 
 In addition, the radiation and matter content of the universe are usually parameterized in terms of the dimensionless numbers, $\orr$ and $\om$, which are defined as
\begin{equation}
 \orr \equiv \frac{8\pi G \rho_R(t_0)}{3 \oh^2}; \quad
\om \equiv \frac{8\pi G \rho_m(t_0)}{3 \oh^2}
\end{equation} 
Here, $t_0$ is the current age of the universe, $H_0\equiv(\dot a/a)_{t_0}$, and $\rho_R(t_0), \rho_m(t_0)$ are the energy densities of radiation and matter in the universe at $t= t_0$. 
[Throughout this review, we use the convention that the symbols $\orr$ and $\om$ always refer to density parameters at the present epoch; when we refer to density parameters, determined at time $t\neq t_0$ (corresponding to the expansion factor $a\neq a_0$), we specifically indicate this dependence in a bracket as $\orr(a)$, $\om(a)$ etc.].
We include all species of particles with the equation of state $p\approx (1/3) \rho$ into radiation and all species of particles with the equation of state $p\approx 0$ into matter. In particular, the matter energy density includes both, that of dark matter particles as well as that of baryons. The interaction between radiation and matter as well as various physical processes in the early universe are irrelevant for the purposes of our discussion and hence we assume that the radiation and matter components  evolve independently.

Thus, we have introduced so far five  parameters ($H_{\rm inf},H_\Lambda, H_0, \orr, \om$) which are mutually independent. The density parameter of dark energy (modeled as a \cc), $\Omega_{\rm DE}$, measured at the present epoch, is given by
$\Omega_{\rm DE} = H_\Lambda^2/H_0^2$ and is \textit{not} an independent parameter. 
It is convenient at this stage to assume that our universe is spatially flat so that $\Omega_{\rm DE} + \orr + \om = 1$.  Since we postulate  the existence of an early inflationary phase of adequate duration in the universe, it is quite reasonable (and may be even mandatory) to assume that the post-inflationary evolution of the universe is described by a spatially flat model.  This condition of spatial flatness,  which is equivalent to  
\begin{equation}
 H_\Lambda^2 = H_0^2 (1-\orr -\om)\, ,
\label{flat}
\end{equation} 
allows us to eliminate  $\oh $ or $H_\Lambda$ in favour of the other.
Thus, the evolution of the spatially flat universe can be described, for example,  in terms of four independent parameters ($H_{\rm inf}, \oh, \orr, \om$).
The  expansion factor $a(t)$  will then be determined by the equation:
\begin{equation}
 \frac{\dot a^2}{a^2} =
 \begin{cases}
 H_0^2 \left[ ( 1 - \Omega_R - \Omega_m)+ \Omega_R a_0^4/a^4 + \Omega_m a_0^3/a^3\right] \qquad\text{(after inflation)}\\
 H_{\rm inf}^2 \qquad \text{(during inflation)}
 \end{cases}
\label{a1}
\end{equation} 
These two equations have to  be matched at the end of inflation, if complications arising from reheating, etc. are ignored.  (We will comment briefly later on about the effects of reheating, etc.) The normalization in \eq{a1} implies that ($\oh, \orr, \om$) are measured when $a = a_0$.  
This equation can be integrated to determine $[a(t)/a_0]$ in terms of the other parameters in the problem. 

We note that the Friedman equations in \eq{s11.3} and \eq{a1} are invariant under the constant rescaling $a\to \lambda a$. Therefore, we \textit{cannot} determine the numerical value of $a_0$ without an extra principle or assumption. This freedom is often used to set $a_0 =1$ which will simplify algebraic expressions. (This issue happens to be more non-trivial than one would have normally assumed, and we make some  comments about this --  as well as an alternative description of Friedmann geometry -- in Appendix \ref{appen:FRWrhop}; but it is not relevant as far as our main discussion is concerned.)

\subsection{Epoch invariant parameterization of cosmology}
\label{sec:epochinv}

The text book parameterization described above  is  convenient for the observational study of the universe because standard observational results are quoted in terms of the parameters ($H_{\rm inf},\oh, \om, \orr$). Theoretically, \eq{a1} describes a four-parameter family of evolutionary histories for different possible universes and our universe is selected out of this infinite set by the specific values for these parameters. At present, these values are thought of as  inputs from observations.

A key agenda item for future research in  cosmology will be to understand why the various parameters which describe our universe have the specific values they have. The above parametrization is useless for this purpose because the parameters ($H_{\rm inf},\oh, \om, \orr$) are tied to the current epoch and have no invariant significance. (For example, any physical theory that predicts, say, $\om=\pi/10$, cannot be taken seriously because this value has no invariant significance!)
To see this more clearly, let us consider cosmologists working at the epoch $t=t_*$ (corresponding to $a(t_*) \equiv a_*$). 
They would have defined the usual quantities $H(t_*)$, the critical density $\rho_c(t_*)\equiv 3 H^2(t_*)/8\pi G$, and the density parameters
$\Omega_m(t_*)\equiv \rho_m (t_*)/ \rho_c(t_*), \ \Omega_R(t_*)\equiv \rho_R(t_*)/\rho_c(t_*)$ and would have written the Friedmann equation (for the post inflationary era) in terms of the three variables $[H(t_*),\Omega_m(t_*),\Omega_R(t_*)]$ as:
\begin{equation}
 \frac{\dot a^2}{a^2}  = \ohs^2\left[[1-\oms-\ors] + \frac{a_*^3\oms}{a^3} + \frac{a_*^4\ors}{a^4}\right]
\label{frw}
\end{equation} 
(The spatial flatness of the universe expressed by \eq{flat} remains valid at all epochs because $H_\Lambda$ is a constant independent of the epoch.)
Thus, the observers working at different epochs, labeled by $t_*$, will use different sets of ``constants'' given by $[H(t_*),\Omega_m(t_*),\Omega_R(t_*)]$ defined with respect to their preferred epoch $t=t_*$. They might also set $a_* =1$ for convenience. Clearly, such a description hides the fact that all these observers are describing the same universe! 

To discuss the cosmic parameters that select our particular universe from the set of all possible universes, we need to rewrite \eq{a1} or \eq{frw} in terms of a set of constants which are epoch independent and have the same value at all times, so that they are \textit{characteristic numbers describing our universe}. In other words, it will be theoretically more useful to describe our universe in terms of certain constants, the
numerical values of which are the same, as measured by any cosmologist working at any epoch. Our first task is to construct such an epoch invariant parameterization of the Friedmann evolution.

 There are several ways to do this  and we will choose a procedure which is most suited for our purpose. 
We begin by noting that, in the total energy density $\rho(a)=\rho_\Lambda+\rho_m(a)+\rho_R(a)$, the part contributed by the
\cc\ [viz., $\rho_\Lambda\equiv 3H_\Lambda^2/8\pi G=\Lambda/8\pi G$] is already epoch-independent. 
Cosmologists working at any epoch will attribute the same numerical value to this quantity. This is also true regarding the constant energy density $\rho_{\rm inf}$ during the inflationary phase (and the corresponding value for $H_{\rm inf}$).
The remaining part $\rho_m(a)+\rho_R(a)$ can be conveniently written as 
\begin{equation}
 \rho_m(a)+\rho_R(a) =\rho_{\rm eq} 
 \left[
 \left(\frac{a_{\rm eq}}{a}\right)^3 +\left(\frac{a_{\rm eq}}{a}\right)^4
 \right]
\end{equation} 
where
\begin{equation}
\rho_{\rm eq}=\textrm{constant}\equiv\frac{\rho_m^4(a)}{\rho_R^3(a)}
\label{defrhoeq}
\end{equation} 
is the density of matter (or radiation) at the epoch $a=a_{\rm eq}$ at which they are equal. From \eq{defrhoeq} it is obvious that numerical value of $\rho_{\rm eq}$ is independent of the epoch $a$ (since $\rho_m(a)\propto a^{-3}$ and $\rho_R(a)\propto a^{-4}$). The quantity $a_{\rm eq}$ is also epoch-independent because it can be defined by the condition:
\begin{equation}
a_{\rm eq}\equiv \frac{a \rho_R(a)}{\rho_m(a)}
 \label{a5}
\end{equation} 
which shows that it is also a characteristic number for our universe.
The density $\rho_{\rm eq}$ allows us to define the temperature $T_{\rm eq} $ of the CMB at the epoch $a=a_{\rm eq}$ by the standard relation $\rho_{\rm eq} = \rho_R(a_{\rm eq}) = (\pi^2/15 ) T_{\rm eq}^4$. Clearly, all cosmologists will attribute the same numerical value to $T_{\rm eq}$ as well.  

The Friedmann equation can now be written entirely in terms of the four epoch-independent parameters $(\rho_{\rm inf}, \rho_\Lambda,\rho_{\rm eq},a_{\rm eq})$ as:
\begin{equation}
 \left(\frac{\dot a}{a}\right)^2 = 
 \begin{cases} 
(8\pi G/3)  \left[\rho_\Lambda + \rho_{\rm eq} (\left(a_{\rm eq}/a\right)^3 +\left(a_{\rm eq}/a \right)^4) \right]\qquad \text{(after inflation)}\\
 8\pi G \rho_{\rm inf} /3  \qquad \text{(during inflation})
 \end{cases}
 \label{aaeq}
\end{equation}
As in the previous case, this equation can be integrated to determine the function
 $x(t)\equiv a(t)/a_{\rm eq}$. It is also clear that $a_{\rm eq} =1$ is an \textit{epoch invariant normalization} of the expansion factor, unlike a choice like $a_0 =1$ or $a_* =1$. All cosmologists working at any epoch can agree to set the expansion factor to unity when the temperature of the CMB had a value $T_{\rm eq}$. It is sometimes convenient, in what follows, to use the variable $H_\Lambda^2=(8\pi G/3)\rho_\Lambda$ and introduce the epoch-independent ratio:
\begin{equation}
 \sigma^4 \equiv\frac{\rho_\Lambda}{\rho_{\rm eq}}=  
 \frac{\rho_R^3}{\rho_m^4} \, \rho_\Lambda  = \frac{\Omega_R^3(t_*)}{\Omega_m^4(t_*)}\, [1-\oms-\ors]
 \label{a7}
\end{equation} 
 in terms of which the Friedmann equation (after inflation) becomes:
 \begin{equation}
 \left(\frac{\dot x}{x}\right)^2 = H_\Lambda^2\left[ 1 + \frac{1}{\sigma^4}\, \left( \frac{1}{x^3} + \frac{1}{x^4}\right)\right]
 \label{a6}
\end{equation}
where $x(t) \equiv a(t)/a_{\rm eq}$ and $x(t) = a(t)$ in the epoch invariant normalization\footnote{It is obvious that any epoch of equality, at which the energy densities of two different components are equal, will allow us to introduce an epoch invariant normalization. But since $\rho_R$ and $\rho_m$ (describing radiation and matter) are better understood theoretically compared to $\rho_\Lambda$, it is prudent to use the equality epoch of matter and radiation densities for our normalization.} $a_{\rm eq} =1$.

Both \eq{a6} and \eq{aaeq} (unlike \eq{frw}) describe the cosmic evolution in terms of epoch-independent parameters. Equation (\ref{a6}) does it in terms of the set $[H_{\rm inf}, \sigma, H_\Lambda]$  while \eq{aaeq} does it in terms of the set $[\rho_{\rm inf}, \rho_{\rm eq}, \rho_\Lambda]$ when we set  $a_{\rm eq}=1$. We will use either set depending on context. 

\subsection{Comparison of $\rho_\Lambda $ and $\rho_{\rm eq}$}

Since the evolution of the universe can be described in terms of three densities $[\rho_{\rm inf}, \rho_{\rm eq}, \rho_\Lambda]$, the set of all universes forms a three parameter family of evolutionary histories. Of these, our universe is selected out by specific values for these three densities. Observationally, we now know that these densities are given by
\begin{eqnarray}
\rho_{\rm inf}&<&(1.94\times 10^{16}\ \text{GeV})^4 \nonumber\\
\rho_{\rm eq} =  \frac{\rho_m^4}{\rho_R^3} &=& [(0.86\pm 0.09) \ \text{eV} ]^4 \nonumber\\
 \rho_\Lambda &=& [(2.26\pm 0.05)\times 10^{-3}  \text{eV}]^4
\end{eqnarray} 
Of these, the post inflationary era of the universe --- which is directly accessible to many different cosmological probes --- is described by the two densities $\rho_{\rm eq}$ and $\rho_\Lambda$. Using natural units $\hbar =c=1$ and the Planck length $L_P= (G\hbar/c^3)^{1/2} = G^{1/2}$, we can construct two dimensionless numbers out of these densities. For the cosmological constant, we get 
\begin{equation}
 \Lambda \left(\frac{G\hbar}{c^3}\right)=8\pi \rho_\Lambda L_P^4\approx 2.8 \times 10^{-122}
\end{equation}
which has led to the cosmological constant ``problem'' for $\Lambda$ since this extremely tiny value for a dimensionless number is supposed to indicate fine-tuning. But, on the other hand, $\rho_{\rm eq}$ leads to the corresponding dimensionless number
\begin{equation}
 (\rho_{\rm eq} L_P^4)\approx 2.4 \times 10^{-113}
\end{equation}
which is hardly commented upon in the literature!
(In fact, it comes as a bit of surprise to many cosmologists that $\rho_{\rm eq} L_P^4$ is indeed such a tiny number!)
As fine-tuning goes, $10^{-113}$ is not much better than $10^{-122}$ for a dimensionless number. Therefore, if we think the value of $\rho_\Lambda L_P^4$ is a ``problem'', then we should also be concerned with the value of $\rho_{\rm eq} L_P^4$.

The conventional wisdom worries about $\rho_\Lambda L_P^4$ but not about $\rho_{\rm eq} L_P^4$ because of two reasons: 
(1) There is a feeling that the cosmological constant is ``somehow'' related to quantum gravity and hence the value $\rho_\Lambda L_P^4$ might have some physical significance.
(2) There is a hope that the numerical value of $\rho_{\rm eq}$ can possibly be determined entirely by high energy physics. From the definition, we can relate $\rho_{\rm eq}$ to the ratio between the number density of the photons and the number density of  matter particles:
\begin{equation}
\rho_{\rm eq}=  \frac{\rho_m^4}{\rho_R^3}
=C \frac{(n_{\rm DM} m_{\rm DM}+n_{\rm B} m_{\rm B})^4}{n_\gamma^4}
=C\left[m_{\rm DM} \left(\frac{n_{\rm DM}}{n_\gamma}\right) + m_{\rm B} \left(\frac{n_{\rm B}}{n_\gamma}\right)\right]^{4}
\label{heprhoeq}
\end{equation} 
where $C =  15^3 (2 \zeta(3))^4  c^3 /\pi^{14} \hbar^3     \approx  2.845 \times 10^{110}$ (in cgs units) is a numerical constant, $n_{\rm DM},n_{\rm B},n_\gamma$ are the current number densities of dark matter particles, baryons and photons respectively and $m_{\rm DM},m_{\rm B}$ are the masses of the dark matter particle and baryon.
 We expect the physics at (possibly) GUTs scale to determine the ratios $(n_{\rm DM}/n_\gamma)$ and $(n_{\rm B}/n_\gamma)$ and specify $m_{\rm DM}$ and $m_{\rm B}$.
 Indeed, we have a framework to calculate these numbers in different models of high energy  physics (for a review, see \cite{mazumdar}) though none of these models can be considered at present as compelling. Since a framework for understanding $\rho_{\rm eq}$ exists, it is not considered to be as mysterious as $\rho_\Lambda$. In fact, the numerical value of $\rho_{\rm inf}$ is also expected to be obtained by high energy physics models, thus leaving $\rho_\Lambda$ as the only constant to be determined. The small value of $\rho_{\rm eq} L_P^4$, or equivalently the small value of 
 \begin{equation}
\Lambda L_P^2 \approx (2.85\pm 0.20) \times 10^{-122} \approx (1.1\pm 0.08)\times e^{-280}.                                                                     \end{equation} 
remains the key problem of which we will be concerned with in this review.

Incidentally, one can rephrase the cosmological constant problem \textit{completely in classical terms} without introducing $\hbar $ or the Planck length $L_P$. We first note that there is \textit{no} fine-tuning problem for the cosmological constant in the classical, \textit{pure gravity} sector. The action for gravity in general relativity is given by
\begin{equation}
 A=\frac{c^4}{16\pi G}\int dt \ d^3 \mathbf{x} \g[R-2\Lambda]
\end{equation} 
which contains three constants, $G, c$ and $\Lambda$. It is not possible to form a dimensionless number from these three constants and hence it is not meaningful to talk about fine-tuning in the context of classical, matter-free gravity. The situation changes when we add matter to the fray. Since the matter sector is characterized by $\rho_{\rm eq}$, we now have the dimensionless constant\footnote{The value of  $\sigma$ can be determined at any epoch and, of course, it is convenient to determine it using the current values of $\om \simeq 0.312 \pm 0.021,\orr = (9.25 \pm 0.38) \times 10^{-5}$  in  \eq{a7}.
The precise values and error bars dependent on the observations one uses and the kind of priors assumed for statistical analysis; for our purpose, the above values, which are consistent with most observations, are adequate.}
\begin{eqnarray}
 \sigma^4 &\equiv&\frac{\rho_\Lambda}{\rho_{\rm eq}}=  
 \frac{\rho_R^3(a)}{\rho_m^4(a)} \, \rho_\Lambda \nonumber\\
&=&[(2.62\pm 0.18) \times 10^{-3}]^4\simeq 10^{-10}\nonumber
\end{eqnarray} 
 The \textit{classical} cosmological constant problem can be stated as the fine tuning of the ratio
\begin{equation}
\frac{\rho_\Lambda}{ \rho_{\rm eq}} \approx 10^{-10} 
\end{equation}
which governs the standard cosmological evolution, structure formation, etc. In fact, all the standard lore in cosmology depends only on this ratio because the post-inflationary evolution of this universe is completely determined by this number. This fine tuning is purely classical and does not require $\hbar$ or $L_P$. Curiously enough, we do not have any direct explanation for the smallness of this number.

To summarize, the family of Friedmann universes we are interested in can be parameterized by a set of three densities $[\rho_{\rm inf},\rho_{\rm eq}, \rho_\Lambda]$ or --- equivalently ---by the three dimensionless numbers
\begin{equation}
 (\rho_{\rm inf} L_P^4, \rho_{\rm eq} L_P^4, \rho_\Lambda L_P^4)\approx(8.3 \times 10^{-17}, 2.4 \times 10^{-113},1.1\times 10^{-123})
\label{sign0}
\end{equation}
Of these, 
we hope that  the
  physics  at, say, the GUTs scale will eventually enable us to estimate the value of $\rho_{\rm inf}L_P^4$ and thus understand the inflationary phase. The value of $\rho_{\rm eq}$, as we said before, can also be determined, in principle, from high energy physics through \eq{heprhoeq}. The worst case, of course, is that of the \cc\ related to the value of  $\rho_\Lambda L_P^4$. 
  In the literature, there have been several attempts to ``solve'' the cosmological constant ``problem'', some of which we rapidly overview in the next section. We then explain why these approaches miss some key ingredients required for a viable solution to the cosmological constant problem. Having done so, we will present a framework which addresses this question fairly comprehensively.

\section{Approaches to the cosmological constant problem}
\label{sec:approaches}

The issues related to the \cc\ have been discussed extensively in the literature and there are several excellent reviews \cite{ccreviews} from different perspectives. The purpose of this short section is to provide a  brief summary of some of these approaches in order to orient the reader toward the  discussions in the later sections of this review.

It is important to realize that the introduction of the \cc\ is closely related to the breaking of a symmetry originally present in the matter Lagrangian. In flat spacetime, the dynamics of matter fields is described by a Lagrangian $L_m (q_A, \partial q_A)$, the variation of which leads to the matter field equations. The addition of a constant ($-\rho$) to $L_m$ in the form $L_m \to L_m -\rho$ is a symmetry of the theory in the sense that the equations of motion for the matter variables $q_A$ remain invariant under this transformation. This continues to be a symmetry even in a \textit{fixed} curved geometry in which the matter Lagrangian could be taken as $\sqrt{-g}\ L_m (q_A, \nabla q_A)$ where $\nabla$ is the covariant derivative; the matter equations of motion still remain invariant under the transformation  $L_m \to L_m -\rho$. The situation changes drastically when we treat gravity, described by the spacetime metric, as a dynamical field and add an action functional for gravity as well, to obtain: 
\begin{equation}
 A_0=\int d^4x \g L_{m}[q_A,\nabla q_A]+\int d^4x \g \left(\frac{R}{16\pi G}\right)
\end{equation} 
In this case, the transformation  $L_m \to L_m -\rho$ leads to the action
\begin{equation}
 A=\int d^4x \g (L_{m}-\rho)+\int d^4x \g\left(\frac{R}{16\pi G}\right)
\label{cstrho}
\end{equation} 
in which the term $[- \g \rho]$ couples the constant to the metric $g_{ab}$ and changes the equations of motion. The action in \eq{cstrho} can be written in the equivalent form :
\begin{equation}
 A=\int d^4x \g L_{m} + \frac{1}{16\pi G}\int d^4x \g \Big[R-2\Lambda\Big]
\end{equation} 
where $\Lambda \equiv 8\pi G \rho$ can be interpreted as the \cc. Obviously, if we add \textit{two} constants to the action, by changing $R \to R - 2\Lambda_1$ and $L_m \to L_m -\rho$, the physics cares only for the total $\Lambda_{\rm tot} = \Lambda_1 + 8\pi G \rho$ and it makes no sense to talk about a separate \cc\ in the gravitational sector \textit{as well as} a shift in the matter Lagrangian. In short, the introduction of a \cc\ can always be thought of as the addition of a constant to the matter sector of the Lagrangian leaving the gravitational action untouched. When the spacetime metric  is treated as a dynamical variable, this breaks a symmetry originally present in the matter sector.

The converse is also true. Any shift in the matter Lagrangian is physically equivalent to the introduction of a \cc. For example, when the universe cools through the energy scale of, say, the electro-weak phase transition, the Higgs potential picks up a large shift in its energy and thus introduces a \cc. The numerical value of this \cc\ is very large compared to what is observed in the universe today thereby  requiring a careful fine-tuning to cancel most of it. This is one of the key problems in understanding the \cc. In the literature, one often comes across emphasis on the divergent contribution to \cc\ from vacuum fluctuations, loop corrections leading to the running of the \cc, etc. While all these effects could contribute to the \cc, it must be stressed that there is a difficulty \textit{even at the tree-level} of the theory  due to the Higgs mechanism operating at, say, the electro-weak state.

The past attempts to tackle the \cc\ problem  were quite diverse. These include (but are certainly not limited to) the following: 
(a) Invoking some symmetry which can be violated by a non-zero \cc\ \cite{symmetrybreak1}, thereby requiring it to be zero. One then needs to break the symmetry weakly to get the observed value of the \cc. 
(b) Making the \cc\ evolve in time, often as a function of an expansion rate: i.e., $\Lambda = \Lambda(H)$. This could be done either in a deterministic fashion (see, for e.g., \cite{ccevol}) or in a stochastic manner \cite{stocc}. 
(c) Using some of the concepts which have been successful in condensed matter physics in related contexts \cite{condmatter}. 
 (d) Attempting to cancel the \cc\ using some back-reaction effect which can act as a self-regulatory mechanism for the \cc. These could arise from different sources including the instabilities in the de Sitter space \cite{backds1,backds2,backds4}.   
(e) Screening of the \cc\ by physical processes \cite{screening}. 
(f) Renormalization group effects leading to the  running of $\Lambda$ \cite{renormgr} finally giving the correct value. 
(g) Introducing non-local or acausal modification of the gravitational action or theories \cite{arkani}.
For example, the \cc\ can be thought of as a Lagrange multiplier for preserving the total four-volume while varying the action, which suggests that the value of the Lagrange multiplier needs to be determined using the four-volume.
(h) Restricting the class of metrics over which the action is varied to get the field equations. For example, if we postulate that $\sqrt{-g}$ should be held fixed while varying the action, $\Lambda$ will decouple from gravity \cite{unimod}.

As we said before, we will not provide a comparative study of the merits and short-comings of all these procedures since such discussions can be found in numerous places in the literature. However, it is fair to say that none of these approaches or their variants have  provided a satisfactory solution to the \cc\ problem. We believe this is because these approaches: (i) address the \cc\ problem somewhat in isolation  and (ii) treat the metric tensor as a dynamical variable describing the gravitational field. We shall now describe what is \textit{really} required to solve the so called \cc\ problem.

\section{Necessary and sufficient conditions to solve the \cc\ problem}\label{sec:nscc}

\hfill {\textit{``It's not that they can't see the solution,}}

\hfill {\textit{it's that they can't see the problem'' }}

\hfill {G.K. Chesterton,}
 
\hfill {\textit{Scandal of Father Brown} (1935)}

\vskip0.2in

\noindent As mentioned before, the \cc\ has a bad press and several ``problems'' have been attributed to it in the literature. A careful scrutiny of these issues reveals that the \cc\ problem(s) essentially arise(s) from two --- and only two --- \textit{distinct}, independent, ingredients which we will describe below. 

Consider any theory of gravity interacting with matter fields and described by a total Lagrangian $L_{\rm tot} = L_{\rm grav} [g_{ab}] + L_m [g_{ab},\phi_A]$. Here, $g_{ab}$ is the metric tensor which is supposed to describe the gravitational degrees of freedom and $\phi_A $ symbolically denotes all other matter degrees of freedom.  Varying the action obtained from  this Lagrangian with respect to the matter degrees of freedom $\phi_A$ will lead to the equations of motion for matter in the presence of a given gravitational field. Similarly, varying the metric tensor will lead to gravitational field equations of the form $\mathcal{G}^a_b = T^a_b$ where $\mathcal{G}_{ab}$ is a geometric variable obtained from the variation of $L_{\rm grav} $ (e.g., it is the Einstein tensor $G_{ab}$ in general relativity, but could be a more complicated tensor in a general theory of gravity like, e.g., in Lovelock models \cite{dktpll}) and $T_{ab}$ is the energy-momentum tensor of matter obtained from the variation of $L_m$ with respect to $g_{ab}$. We also note, for future reference, that the scalar nature of the Lagrangians lead to the generalized  Bianchi identity
($\nabla_a \mathcal{G}^a_b =0$)
and the conservation of the energy-momentum tensor $(\nabla_a T^a_b =0)$ in any theory.

These variational principles leading to the equations of motion for matter and gravity, contain an important peculiarity.  The equations of motion for the matter sector are invariant under the addition of a constant to the matter Lagrangian. That is, the transformation
\begin{equation}
 L_m\to L_m + ({\rm constant})
 \label{lmsym}
\end{equation} 
is a symmetry of the matter sector in the sense that the matter equations of motion remain invariant under the transformation in \eq{lmsym}. (We will assume that the matter Lagrangian is \textit{not} supersymmetric invariant in the 
regime we are interested in; supersymmetry  is the only symmetry that prevents the addition of a constant to the Lagrangian.)
Though the addition of a constant to a Lagrangian is usually not thought of as a ``symmetry'' principle, it definitely should be, because this mathematical operation does leave the equations of motion invariant; in that sense, it is no different from any other mathematical operation, like for e.g., the Lorentz transformations which leave the equations of motion invariant. Usually, one constructs the Lagrangian (or, more precisely, the action) in such a way that it itself remains manifestly invariant under all the relevant symmetry transformations (like, for e.g., the  Lorentz transformation). But, in the case of \eq{lmsym}, the Lagrangian and action do change but the matter equations of motion 
do not. 
In fact, if we adopt the principle that Lagrangians should be written in a manner making all the symmetries apparent, then the matter sector Lagrangian should always be expressed in the form $(L_m + C)$ 
where $C$ is left as an unspecified constant. (This is analogous to not choosing a specific Lorentz frame while writing a Lagrangian, thereby exhibiting manifest Lorentz invariance.)\footnote{As an aside, we make the following comment: In the context of non-gravitational physics (classical mechanics, quantum mechanics,.....) we do know that it is only the differences in the energy which matter, and not the absolute zero of the energy. Nevertheless, it surprising that \textit{no} formalism of non-gravitational physics, say, elementary classical or quantum mechanics, exists, which deals \textit{directly} with energy differences! The natural theoretical development of these theories uses energy itself --- not energy differences. This is somewhat analogous to the natural description of gauge field theory, like electrodynamics, which uses the gauge potential ($A_k$) though the observables are built from the field strengths $F_{ik}$ (in the classical context) and line integrals over gauge potentials (in the quantum theory). There is an approach, usually called relational dynamics, which attempts to study interacting particles using only relative velocities, differences in co-ordinates, etc. Even in this context, we are not aware of a formulation which uses only the differences in energy rather than the energy itself. Given this background, it is indeed interesting that there exists a formulation of gravitational theories --- as we shall see later --- which is immune to the absolute value of the energy.}
Keeping such an arbitrary constant $C$ in the matter sector allows one to make manifest that the equations of motion for matter are indeed invariant under \eq{lmsym} because any additional constant merely changes the value of $C$.

However, the gravitational field equations, in contrast to the matter field equations, are \textit{not} invariant under the addition of a constant to the matter Lagrangian.
The energy-momentum tensor of the matter is changed by the addition of a constant, as: 
\begin{equation}
 T^a_b \to T^a_b + ({\rm constant}) \ \delta^a_b
 \label{tsym}
\end{equation} 
Hence, the gravitational field equations now become $\mathcal{G}^a_b = T^a_b + ({\rm constant}) \ \delta^a_b$. This is equivalent to the introduction of a \cc, if one was not present originally, or a change in its numerical value, if a \cc\ was present in the gravitational Lagrangian. 

This is the crucial problem related to the \cc\ viz., that its numerical value (either zero or non-zero) can be altered by the transformation in \eq{lmsym}, which, however, leaves the matter equations unchanged. In other words, particle physicists interested in the standard model may choose the overall constant in the matter Lagrangian arbitrarily since the standard model is unaffected by this constant. However, each choice for the constant leads to a different value for the cosmological constant, and, hence, a different geometry for the universe. Many of these, of course, are observationally untenable.

We may restate the above problem as follows: If a fundamental principle that enables the determination of the numerical value of the cosmological constant (either zero or non-zero) is discovered, it is not useful if the gravitational field equations are not invariant under the transformations in \eq{lmsym} or \eq{tsym}.  Note that:

\begin{itemize}
 \item 
This problem  is quite fundamental and is independent of the actual value of the cosmological constant determined by observations (either zero or non-zero). The problem of the \cc\ existed even before observations showed that it might have a non-zero value!
\item
The problem is also independent of issues related to the energy densities of vacuum \textit{fluctuations}, regularization of zero-point energies, etc. If the Higgs mechanism operated during the evolution of the universe, causing the zero level of the energy densities to change by a large factor, then we face a \cc\ problem in the form of extreme fine-tuning already \textit{at the tree-level} of quantum field theory.
\end{itemize}

From the above discussion, we can immediately identify three ingredients which are \textit{necessary} to solve the \cc\ problem: 
\begin{enumerate}
 \item The field equations of gravity \textit{must be} made invariant under the transformations in \eq{lmsym} and \eq{tsym} so that gravity is ``immune'' to the shift in the zero level of the energy densities. \label{page:conditions}
 
 \item At the same time, however, the solutions to the field equations \textit{must} allow the cosmological constant to influence the geometry of the universe, because without this, the observed accelerated expansion of the universe  cannot possibly be explained. 
 
 \item Since the \cc\ cannot be introduced as a low energy parameter in the Lagrangian if the theory is invariant under the transformation in \eq{lmsym}, we need a fundamental physical principle to determine its numerical value.
\end{enumerate}

At first sight, requirements (1) and (2) might sound impossible to satisfy simultaneously. However, it can be achieved by constructing a set of gravitational field equations which are invariant under the transformation in  \eq{tsym}  but allow the inclusion of a \cc\
as an \textit{integration constant} in their solutions. For example, consider a theory in which the field equations are given by the requirement:
\begin{equation}
 (\mathcal{G}^a_b - T^a_b) n_a n^b = 0
 \label{tf}
\end{equation} 
for all \textit{null} vectors $n^a$ in the spacetime \cite{aseemtp}. The above equations can be solved  by $\mathcal{G}^a_b - T^a_b = F(x) \delta^a_b$, but the generalized  Bianchi identity
($\nabla_a \mathcal{G}^a_b =0$)
and the conservation of the energy-momentum tensor $(\nabla_a T^a_b =0)$ together imply that $F(x)$ is a constant. Hence, \eq{tf} is equivalent to the standard gravitational field equations with an arbitrary \cc\ appearing as an integration constant. Thus, if we can construct a theory of gravity in which the field equations reduce to those in \eq{tf}, then the  first two requirements for solving the \cc\ problem would have been achieved. 

The above demand turns out to be extremely strong, and it has important consequences which are often overlooked in attempts to ``solve'' the \cc\
problem. To see this, consider any theory of gravity interacting with matter, which satisfies the following three conditions: 
\begin{enumerate}
 \item The theory is generally covariant, and hence the matter action is constructed by integrating  a scalar Lagrangian $L_m(g_{ab},\phi_A)$ over the measure $\sqrt{-g} d^4x$. 
 \item The equations of motion for matter are invariant under the transformation $L\to L + C$, where $C$ is a scalar constant. \label{page:conditions1}
 \item The field equations for gravity are obtained by an unrestricted variation of the metric tensor $g_{ab}$ in the total action.
\end{enumerate}
It is easy to see that the \cc\ problem  \textit{cannot} be solved in any theory satisfying the above three requirements. (This was also emphasized clearly in Section IV of ref. \cite{tpap}). In particular, we cannot obtain the gravitational field equations of the form in \eq{tf} in any theory that satisfies the above criteria.

Even though all the three criteria stated above seem reasonable, together, they prevent us from solving the \cc\ problem. Hence, one of them needs to be given up; assuming we do not give up general covariance of the theory, or the freedom to add a constant to the matter Lagrangian, we may only modify the third requirement. 

A simple way of obtaining \eq{tf} is the postulate that the gravitational field equations are obtained by varying the metric, but keeping $\sqrt{-g} = $ constant. Such theories, known as unimodular theories of gravity, have been studied in the literature in the past \cite{unimod}. Unfortunately, there is, at best, weak or, at worst, non-existent motivation to keep $\sqrt{-g} = $ constant.  

It is also possible to obtain \eq{tf} from the alternative perspective of gravity that treats gravity as an emergent phenomenon (see ref.\cite{aseemtp}; for a review, see \cite{tpreviews}). In this approach, we associate thermodynamic potentials with all null vector fields in a spacetime. The maximization of the relevant thermodynamic potential (like entropy, free energy, ...) associated with all null vectors simultaneously, then leads to \eq{tf}. Maximization involves varying the null vector fields, rather than the metric, and hence it sidesteps the difficulties by dropping the third requirement in the above list.
The metric is not varied to obtain the field equations.  In  this approach, even the original action principle is invariant under the transformation in \eq{tsym}. (We shall describe some of these results in Sec.\ref{sec:egandcc}).

But, as pointed out earlier, constructing a theory which is invariant under \eq{tsym} is only winning half the battle. The resulting field equations in \eq{tf} allow an arbitrary integration constant in the solution which has to be fixed once (and only once) for our universe.  But its numerical value cannot be determined from the theory or from parameters of the bulk Lagrangian because the gravitational field equations are immune to them. It follows that we need some other extra physical principle to determine the numerical value of this integration constant \cite{tpdetcc}.

In the next section (Sec.\ref{sec:cosmin}), we shall relate the cosmological constant to a new conserved quantity (which we called \cn) present in our universe and reduce the problem of determining the value of $\Lambda L_P^2$ to that of determining the numerical value of \cn. Both observational and theoretical considerations suggest that the value of \cn\ is $4\pi$, which, in turn, allows us to relate $\rho_\Lambda$ to $\rho_{\rm eq}$ and $\rho_{\rm inf}$. This new physical principle (viz. \cn\ = $4\pi$) in the context of the emergent gravity paradigm, provides a comprehensive solution to the cosmological constant problem.
 In Sec.\ref{sec:egandcc}, we describe in detail how these ideas fit into the broader context of cosmological evolution in the emergent gravity paradigm.

\section{CosMIn and the solution to the cosmological constant problem}
\label{sec:cosmin}

We shall now describe a physical principle capable of determining the value of the \cc. 
 It is obvious that $\ln (\Lambda L_P^2)$ is a more tractable quantity than $\Lambda L_P^2$ itself.  If $\ln (\Lambda L_P^2)$ can be related  to a physically meaningful parameter, the value of which can be independently understood, then we obtain a handle on the numerical value of the cosmological constant. We will introduce a number $N_c$ --- which we call the `Cosmic Mode Index', or \cn, for short ---  that is related to $\ln (\Lambda L_P^2)$ and counts the number of modes within the Hubble volume that cross the Hubble radius in the radiation and matter dominated eras --- that is, during the period between the end of inflation and the beginning of late-time acceleration \cite{gr2013}. The value of \cn\ \textit{is a characteristic number of our universe} and we will express this number as a function of the epoch-invariant parameters describing the universe, i.e. $[\rho_{\rm inf},\rho_{\rm eq},\rho_\Lambda]$
or $(\rho_{\rm inf} L_P^4, \sigma, \Lambda L_P^2)$. We do this in Sec.\ref{sec:incn};
having done this, we have two ways of proceeding: 

(a) A rather conservative approach is to determine the value of \cn\ from the observationally determined values of, say, $[\rho_{\rm inf},\rho_{\rm eq},\rho_\Lambda]$ and try to understand this value (Sec.\ref{sec:cnfrobs}). Looking ahead, we mention that observations  lead to the remarkable result that $N_c\approx 4\pi$; or more precisely, $N_c= 4\pi \mu$ with $\mu \approx 1.00\pm 0.001$. \textit{Discovery of this numerical result, by itself, is a  finding of some importance.} \label{page:list}

(b) A more fundamental approach is to postulate a value for \cn\ from theoretical motivations, thereby relating the three epoch invariant parameters, say $[\rho_{\rm inf},\rho_{\rm eq},\rho_\Lambda]$. Then, given the value of any two of them, we can compute the third. As we said before, high energy physics is (in principle) capable of determining the values of $[\rho_{\rm inf},\rho_{\rm eq}]$. Therefore using our relation, we can determine the value of $\rho_\Lambda=f(\rho_{\rm inf},\rho_{\rm eq})$ in terms of the other two densities (Sec.\ref{sec:ncc}).

We shall describe both these approaches, though our preference is for approach (b).

\subsection{Introduction of \cn}\label{sec:incn}

To motivate the definition of \cn, we proceed as follows:
The Hubble radius $H^{-1}(a) \equiv (a/\dot a)$ remains constant during the inflationary and the late-time accelerating phases, while it evolves as $H^{-1}(a) \propto a^2$ in the radiation dominated phase and as $H^{-1}(a) \propto a^{3/2}$ in the matter dominated phase. An important concept in standard cosmological models is the crossing of the Hubble radius by proper length scales characterized by a co-moving wave number, $k$, related to the proper length $\lambda_{\rm prop}(a) \equiv a/k$.   This crossing  occurs whenever the equation $\lambda_{\rm prop}(a)=H^{-1}(a)$, i.e., $k=a H(a) $ is satisfied. 

\begin{figure}
 \begin{center}
  \includegraphics[scale=0.35]{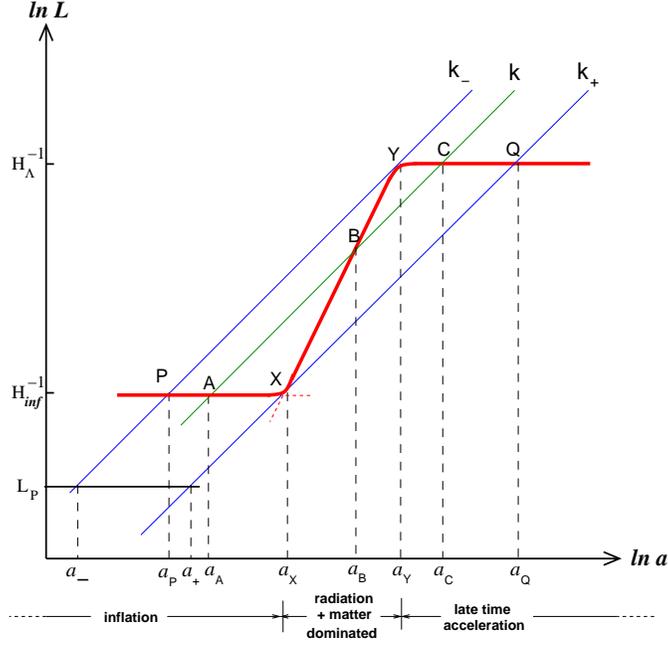}
 \end{center}
\caption{
The Hubble radius $H^{-1}(a)$ of the universe (thick red line $PAXBYCQ$) delineates the three phases of evolution of our universe: (i) inflationary phase ($a<a_X, H^{-1}=H^{-1}_{\rm inf}=$ constant); (ii) radiation/matter dominated phase ($a_X<a<a_Y, H^{-1}\propto a^2$ or $a^{3/2}$); (iii) late-time accelerated phase ($a_Y<a, H^{-1}=H^{-1}_{\Lambda}=$ constant). A generic length scale parametrized by a comoving wavenumber $k$ and proper length $\lambda_{\rm{prop}}(a) \equiv a/k\propto a$ (line $ABC$) exits the Hubble radius during inflation (A), re-enters during the radiation/matter dominated phase (B) and again exits during late-time acceleration (C). But  modes with $k< k_-$ (line $YP$ corresponds to $k=k_-$) or   with $k> k_+$ (line $XQ$ corresponds to $k=k_+$) cross the Hubble radius only once. The points $P$ (obtained by drawing a unit slope tangent at $Y$ and extending it backwards) and $Q$ (obtained by drawing a unit slope line at $X$ and extending it forwards) determine the cosmic parallelogram $PXQY$ of the relevant length scales \cite{bj,tpbj1,tpbj2}. The number of length scales which cross the Hubble radius between $X$ and $Y$ (which is the same as the number of length scales that cross during $PX$ or $YQ$) is a unique number $N_c$ for our universe. Observations show that it is very close to $4\pi$. Note that these length scales, when extrapolated back, cross the Planck length $L_P$ during the interval $a_-<a<a_+$.
}
\label{fig:hubble}
\end{figure}

For a \textit{generic} length scale (see Fig.\ref{fig:hubble}; line marked $ABC$), this equation has three solutions: first at $a=a_A$ (during the inflationary phase; point $A$ in Fig.\ref{fig:hubble}), second, at $a=a_B$ (during the radiation/matter dominated phase; point $B$) and the third one at $a=a_C$ (during the late-time accelerating phase; point $C$).
But note that length scales with $k<k_-$ exit during the inflationary phase and never re-enter, while length scales with $k>k_+$ remain inside the Hubble radius until very late and only exit during the late-time accelerating phase. 

It is now obvious that all the length scales which leave the Hubble radius during the interval $PX$ (in \fig{fig:hubble}) will enter the Hubble radius during the interval $XY$ in the radiation/matter dominated phase and will again leave the Hubble radius during the interval $YQ$ during the late time accelerated phase. \textit{This fact provides a deep and fundamental link between the three phases of evolution of the universe which are otherwise totally unconnected and specified by rather ad-hoc parameters in the conventional description.} If we can count the number of proper length scales which leave the Hubble radius during $PX$, re-enter during $XY$ and leave again during $YQ$, then such a number will be a clear signature of our universe. What is more, that number will also provide a link between the three phases since it remains constant during the three different crossings of the Hubble radius by the relevant length scales.

To determine the number of proper length scales involved in this process, we need to introduce a  suitable measure and \textit{define} a quantity $N(a_1, a_2)$ which counts the number of length scales which cross the Hubble radius during an arbitrary interval $(a_1<a<a_2)$. A natural measure which can be used to define this quantity can be arrived at as follows. If we think of the length scales as the modes of some field carrying a single degree of freedom, then we know that the modes with comoving wave numbers in the interval $(k, k+dk)$ where $k=aH(a)$ and $dk=[d(aH)/da] da$ will cross the Hubble radius during the interval $(a,a+da)$. The number of such modes in a comoving Hubble volume $V_{\rm com}=(4\pi H^{-3}/3a^3)$ with wave numbers in the range  $(k, k+dk)$ is $dN = V_{\rm com} d^3k/(2\pi)^3 $. Hence, we \textit{define} the number of length scales that cross the Hubble radius during $a_1 < a < a_2$ to be:
\begin{eqnarray}
 N(a_1,a_2) &=& \int_{a_1}^{a_2} \frac{V_{\rm com} k^2}{2\pi^2} \, \frac{dk}{da}\ da
= \frac{2}{3\pi}\int_{a_1}^{a_2} \frac{d(Ha)}{Ha} \nonumber \\
 &=& \frac{2}{3\pi}\ln \left(\frac{H_2a_2}{H_1a_1}\right),
\label{Ndef}
\end{eqnarray} 
where we have used the relations $V_{\rm{com}}=4\pi/3H^3a^3$ and $k=Ha$. 
We shall take $N(a_1,a_2)$ given by the above expression as a natural measure of the number of proper length scales which cross the Hubble radius during the interval $(a_1<a<a_2)$. 

To avoid possible misunderstanding, we stress that we are \textit{not} thinking at this stage in terms of any specific field modes (like in the case of inflationary perturbation theory, etc.) but merely use the correspondence with that context in order to \textit{define }$N(a_1,a_2)$. This should be treated purely as a \textit{definition} which counts the number of proper length scales which cross the Hubble radius during a specified interval. Of course, it will \textit{also} be numerically equal to the number of modes per unit spin degree of freedom of any  field which is present in the universe but that feature is irrelevant for our discussion at present.
In the above definition we have made natural choices for some numerical factors, which, as we shall see,  lead to interesting and acceptable results. Also note that $r(t)/a(t)=1/k=$ constant, where $r$ is the proper distance from some origin, describes geodesics in the Friedmann universe. Thus, every wave number $k$ labels a radial geodesic in the spacetime. So, the length scales which we are interested in are also in one-to-one correspondence with the geodesics in the spacetime. 

The number $N$ has several simple properties. It is well defined for any range $(a_1 < a < a_2)$ and is  invariant under multiplication of $a$ by an arbitrary factor. (As a consequence, it is independent of $a_{\rm eq}$.) The way it is defined, $N$ is positive when $H_2 a_2 > H_1 a_1$ and negative otherwise. But very often we are only interested in the magnitude of $N$, and we choose the sign suitably to keep it positive. Further, if $a_1$ and $a_2$ are chosen to be epochs at which a given mode with wave number $k$  crosses the Hubble radius, so that $H_1a_1 = H_2 a_2 = k$, then $N( a_1, a_2)=0 $.  For the generic mode in Fig.\ref{fig:hubble} which crosses the Hubble radius thrice, at $A$, $B$ and $C$, the value of $N(a_A, a)$ increases from $a=a_A$ till the end of inflation at $a=a_X$, and then decreases from $a=a_X$ to $a=a_B$, reaching zero again at $a=a_B$. 
Every length scale which exits the Hubble radius during the interval $a_A<a<a_X$ re-enters the Hubble radius during the interval $a_X<a<a_B$. 

As we said earlier,   the number of modes $N(a_X, a_Y)$ which enter the Hubble radius during the radiation/matter dominated era is a \textit{characteristic number for our universe}, which we call \cn\ and denote its \textit{magnitude} (disregarding the sign) simply as $N_c$.  \cn\ counts these modes which exit the Hubble radius during $a_P<a<a_X$ in the inflationary phase, re-enter during $a_X<a<a_Y$ in the intermediate phase and again exit during $a_Y<a<a_Q$ in the late-time accelerating phase. It is, in fact,  possible to argue  \cite{bj,tpbj1,tpbj2} that the cosmologically relevant part of the evolution is located inside the cosmic parallelogram $PXQY$.
Obviously, it is of interest to compute this number $N_c$ since it is a clear characteristic for our universe and is a common feature connecting the three phases of evolution.

Computing  $N_c$ for our universe is straightforward and  involves the following  steps: 
 We first determine $x_2,x_1$ corresponding to the epochs $a_Y,a_X$ in Fig.\ref{fig:hubble} using \eq{a6}. The epoch $x_2=x_2(H_\Lambda,\sigma)$  is determined (as a function of $H_\Lambda,\sigma$) by the condition that the tangent to the curve has unit slope, which is equivalent to $d[aH(a)]/da=0$.
It is straightforward to show from this, that $x_2$ satisfies the quartic equation $\sigma^4 x_2^4 = (1/2) x_2 + 1$. To determine $x_1$, we need to connect the Hubble radius after inflationary reheating with its value during the inflationary phase, $H_{\rm inf}$. This will depend on the detailed modeling of the reheating,  but, if we assume that all the energy in the inflation field was instantaneously converted into radiation (efficient reheating) then  $x_1$
can be determined by matching  $H(a) $  in the radiation dominated phase to the Hubble constant during the inflation, $H_{\rm inf}$, giving  $x_1 \simeq H_{\rm inf}^{-1/2}$. If the reheating was not efficient, this expression -- and all or further results -- still hold with $H_{\rm inf}$ being interpreted as $H_{\rm reheat}$, viz. the value of $H$ after reheating. Similarly, $\rho_{\rm inf}$ should be interpreted as $\rho_{\rm reheat}$ viz. the energy density at the end of reheating. (Generically $H_{\rm inf}>H_{\rm reheat}$ and $\rho_{\rm inf}>\rho_{\rm reheat}$ because the reheating may not be efficient; we will comment on the effect of reheating whenever relevant.)
 Once we know $x_1(H_{\rm inf})$ and $x_2(H_\Lambda,\sigma)$, we can  express $N_c$ in \eq{Ndef}  as a function of $H_\Lambda, \sigma, H_{\rm inf}$. 
We will now describe the results obtained by the above procedure; the calculational details are given in Appendix \ref{appen:cncalc}. 

Following the above procedure, we find that $\mu\equiv N_c/4\pi$ can be expressed in terms of $(\rho_{\rm inf} L_P^4, \sigma, \Lambda L_P^2)$ in the form:
\begin{equation}
\mu=\frac{N_c}{4\pi}=\frac{1}{24 \pi^2} \ln \left(\frac{8 \pi C(\sigma) \rho_{\rm inf} L_P^4}{3 \Lambda L_P^2} \right)
\label{eqformu}
\end{equation} 
where $C(\sigma) = 6 (r+2) (3 r + 4)^{-2}$ and $r(\sigma)$ satisfies the quartic equation $\sigma^4 r^4 = (1/2) r + 1$. Again, this result is valid for efficient reheating. More generally, it holds with $\rho_{\rm inf}$ interpreted as $\rho_{\rm reheat}$, viz. the energy density at the end of reheating. In what follows, we shall continue to use the symbol $\rho_{\rm inf}$ with this understanding.

\subsection{Determining \cn\ from observations}\label{sec:cnfrobs}
As mentioned earlier, we now have two ways of proceeding, indicated by (a) and (b) in page \pageref{page:list}. We shall first proceed along the lines of (a).

\begin{figure}
 \begin{center}
  \includegraphics[scale=0.9]{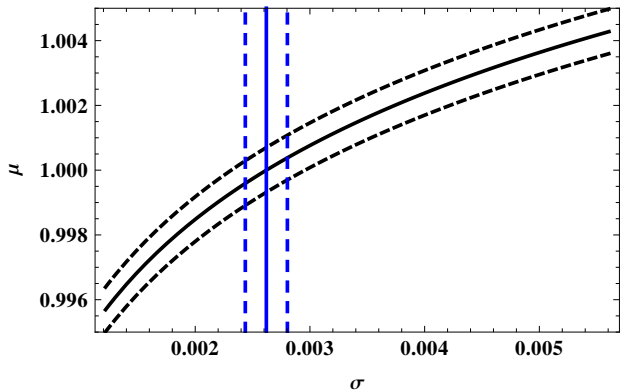}
 \end{center}
\caption{This figure shows the range of values of $\mu \equiv N_c/4 \pi$ for the range of $\sigma$ allowed by cosmological observations, i.e. $\sigma = (2.437 - 2.803) \times 10^{-3}$. The range in $\sigma$ is indicated by the vertical dashed blue lines. The fiducial value of $\sigma$ is $2.62 \times 10^{-3}$, indicated by the solid vertical blue line. The solid black curve shows the variation of $\mu$ with $\sigma$ for the fiducial value of the energy scale of inflation  $\rho_{\rm inf}^{1/4}=1.166 \times 10^{15}$ GeV. The dashed black curves show the variation of $\mu$ with $\sigma$ for the values of the energy scale of inflation in the range $(1.120-1.216) \times 10^{15}$ GeV if the reheating is fully efficient. (If not, this is a lower bound and the actual $\rho_{\rm inf}^{1/4}$ may be about a factor 10-15 larger.) It can be seen that the range of $\mu$ allowed by this variation is of the order of a few parts in thousand.}
\label{fig:muwithcosmo}
\end{figure}

\begin{figure}
 \begin{center}
  \includegraphics[scale=0.9]{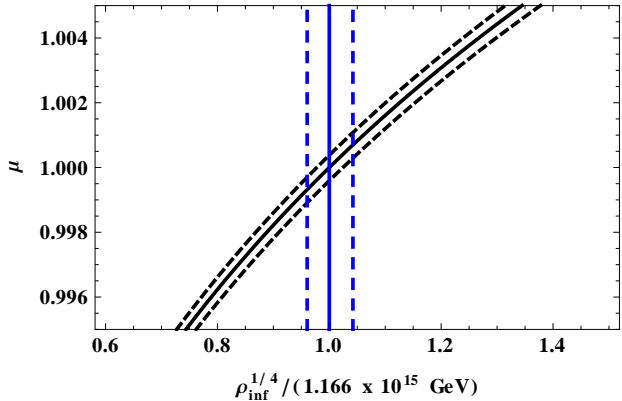}
\end{center}
\caption{This figure shows the range of values of $\mu \equiv N_c/4 \pi$ for the energy scale of inflation in the range $\rho_{\rm inf}^{1/4}=(1.120-1.216) \times 10^{15}$ GeV indicated by the vertical dashed blue lines. (If the reheating is not efficient, this represents $\rho_{\rm reheat}^{1/4}$ and the actual $\rho_{\rm inf}^{1/4}$ can be a factor 10-15 larger. We shall use continue to use the symbol $\rho_{\rm inf}$ with this understanding.) The fiducial value of the energy scale of inflation is $\rho_{\rm inf}^{1/4}=1.166 \times 10^{15}$ GeV, indicated by the solid vertical blue line.
The solid black curve shows the variation of $\mu$ with $\rho_{\rm inf}$ for the fiducial value of $\sigma = 2.62 \times 10^{-3}$, as indicated by observations. The dashed black curves show the variation of $\mu$ with $\rho_{\rm inf}$ for the values of $\sigma = 2.437 \times 10^{-3}$ and $2.803 \times 10^{-3}$, encompassing the allowed range of $\sigma$ permitted by cosmological observations. It can be seen that the range of $\mu$ allowed by this variation is of the order of a few parts in thousand.}
\label{fig:muwithbeta}
\end{figure}

We can determine the numerical value of \cn\ from the observationally determined values of $(\rho_{\rm inf} L_P^4, \sigma, \Lambda L_P^2)$. For a GUTs scale inflation with $\rho_{\rm inf} = (1.2\times 10^{15}$ GeV)$^4$, $\sigma = 2.62\times 10^{-3}$ and $\Lambda L_P^2 = 2.85 \times 10^{-122}$ determined by cosmological observations, we get $\mu=1.0$ !
Taking into account the uncertainties in cosmological observations, $\sigma$ is in the range $(2.437 - 2.803) \times 10^{-3}$ which translates to \begin{equation}
\mu = 1.0000 \pm 0.0004.
\label{muvalue}                                                                                                                                                                                     \end{equation} 
Similarly, if we vary the energy scale of inflation in the range $1.120 \times 10^{15} \ \rm{GeV} <\rho_{\rm inf}^{1/4}< 1.216 \times 10^{15} \ \rm{GeV}$, and keep $\sigma=2.62 \times 10^{-3}$ fixed, we get 
\begin{equation}
\mu=1.0000 \pm 0.0007.                      
\end{equation} 
These results are shown in Figures \ref{fig:muwithcosmo} and \ref{fig:muwithbeta}. In \eq{eqformu}, because of the factor $24\pi^2$ in the denominator and the logarithmic dependence, $\mu$ varies slowly with respect to the parameters. For example, if the inflationary energy scale $\rho_{\rm inf}$ changes by a factor 5, $\mu$ will change only by
$(24\pi^2)^{-1}\ln 5\approx 0.007$. \textit{It is clear that $\mu$ is equal to unity to a high order of  accuracy} for an acceptable range of cosmological parameters. 

As we saw earlier, the evolution of the universe can be described in terms of three epoch-independent parameters which could be taken conveniently as either one of the two sets of numbers: $(\rho_{\rm inf} L_P^4, \sigma,  \Lambda L_P^2)$ or  $(\rho_{\rm inf}L_P^4, \rho_{\rm eq}L_P^4,\rho_\Lambda L_P^4)$. We also found that the numerical values of these parameters --- especially $\Lambda L_P^2$ --- 
are not easy to interpret.  We have now defined another parameter, \cn, which is yet another characteristic number for our universe, in terms of the three original variables. This, by itself, allows us to trade off one of the three variables in the sets of epoch-independent parameters  in favour of $N_c$.
For example, in the set $(\rho_{\rm inf} L_P^4, \sigma, \Lambda L_P^2)$, we can eliminate $\Lambda L_P^2$ in terms of $\sigma, \rho_{\rm inf} L_P^4$ and $N_c$ using the relation
\begin{equation}
 \Lambda L_P^2 = \frac{8 \pi \rho_{\rm inf} L_P^4 C(\sigma)}{3}\exp[-6\pi N_c]
\label{holy3}
\end{equation} 
In other words, we could have equivalently chosen to describe the universe in terms of the new set of epoch-independent parameters
\begin{equation}
 (\rho_{\rm inf} L_P^4, \sigma, \Lambda L_P^2)\rightarrow (\rho_{\rm inf} L_P^4, \sigma, N_c)
 \approx 
(\rho_{\rm inf} L_P^4, \sigma, 4\pi)
\label{sign1}
\end{equation} 
where at least one parameter seems to have an interesting ($4\pi$) numerical value for our universe, among all possible universes.

In this parameterization, we are assuming that $\sigma$ is specified in terms of $\orr$ and $\om$ which is convenient for comparison with observations. But from a more fundamental point of view, it is $\rho_{\rm eq} L_P^4$ which is determined by high energy physics. It is also possible to eliminate $\rho_\Lambda L_P^4$ in terms of $(\rho_{\rm inf} L_P^4, \rho_{\rm eq} L_P^4,N_c)$ by rewriting $\sigma$ as $\rho_\Lambda/\rho_{\rm eq}$ in \eq{holy3} and using $\Lambda L_P^2 = 8\pi(\rho_\Lambda L_P^4)$.  This expression is algebraically a bit more complicated because determining $C(\sigma)$ requires solving a quartic equation. However, for all practical purposes (when $\sigma^4\ll 1$), a good approximation to $C(\sigma)$ is given by $C(\sigma) \approx (1/3) (2\sigma)^{4/3}$.
Using this, it is easy to obtain the equivalent of \eq{holy3} in the form:
\begin{equation}
 \rho_\Lambda  = \frac{4}{27} \frac{\rho_{\rm inf}^{3/2}}{\rho_{\rm eq}^{1/2}} \exp (- 9\pi N_c)
\label{ll4}
 \end{equation} 
We can now work with the set of parameters 
\begin{equation}
 (\rho_{\rm inf}L_P^4, \rho_{\rm eq}L_P^4, \rho_\Lambda L_P^4) \rightarrow (\rho_{\rm inf}L_P^4, \rho_{\rm eq}L_P^4, N_c)
 \rightarrow (\rho_{\rm inf}L_P^4,\rho_{\rm eq}L_P^4, 4\pi)
 \label{sign2.5}
\end{equation} 
We now see that the result $N_c\approx 4\pi$ has eliminated one of the  parameters, viz. the \cc,  while the two parameters $\rho_{\rm eq}L_P^4$ and $\rho_{\rm inf}L_P^4$ will be determined by high energy physics. 

\subsection{Numerical value of the cosmological constant}\label{sec:ncc}

We now consider the approach (b) mentioned on page \pageref{page:list}. To begin with, we elevate the observational result $N_c=4\pi$ to the status of a postulate. 
 This postulate allows us to determine the numerical value of $\Lambda L_P^2$. 
With the postulate that $N_c=4\pi$, we can write \eq{holy3} as
\begin{equation}
 \Lambda L_P^2 = \frac{8 \pi \rho_{\rm inf} L_P^4 C(\sigma)}{3}\exp[-24\pi^2]
\label{holy2}
\end{equation} 
where $C(\sigma) = 6 (r+2) (3r+4)^{-2}$ and $r$ satisfies the quartic equation $\sigma^4 r^4 =   (1/2) r + 1 $. 
Given the value of $\sigma$ from observations (through $\om$ and $\orr$) and the 
inflation scale determined by $\rho_{\rm inf} L_P^4$,  we can \textit{calculate} the value of the cosmological constant from \eq{holy2}. For a GUTs scale inflation with $E_{\rm inf} = 1.2\times 10^{15}$ GeV and $\sigma = 2.6\times 10^{-3}$ (determined by cosmological observations), we get $\Lambda L_P^2 = 2.853 \times 10^{-122}$ which agrees with the observed value. The bulk of this ``smallness'' is contributed by the $\exp(-24\pi^2)$ factor in \eq{holy2}.

\begin{figure}
 \begin{center}
  \includegraphics[scale=0.85]{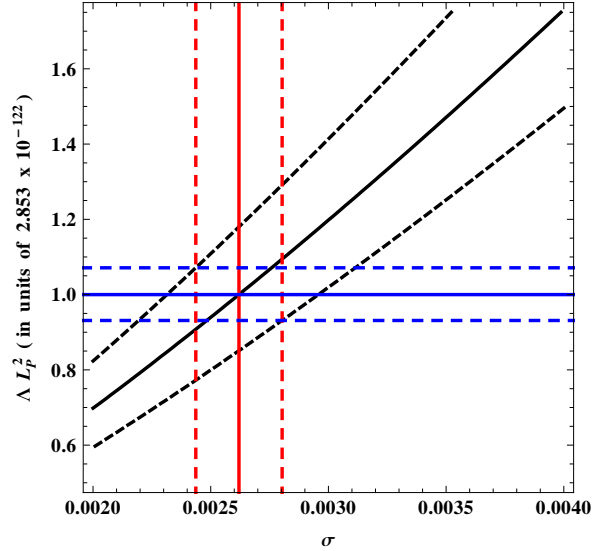}
 \end{center}
\caption{The postulate that $N_c=4\pi$   determines the numerical value of $\Lambda L_P^2$ in terms of (i) the epoch-invariant parameter
$\sigma \equiv(\Omega_R^{3/4}/\Omega_m)\, (1-\Omega_m - \Omega_R)^{1/4}$  and (ii) the  inflationary scale parametrized by $\rho_{\rm inf} L_P^4$. Observational constraints \cite{planck2013,cosmoreview} suggest $\sigma = (2.437 - 2.803) \times 10^{-3}$, which is shown by the three vertical (red) lines. The thick black curve is obtained from \eq{holy2} if we consider the inflationary energy scale of $\rho_{\rm inf}^{1/4}=1.166 \times 10^{15}$ GeV, and leads to the observed value of $\Lambda L_P^2=2.835 \times 10^{-122}$, shown by the horizontal unbroken (blue) line. The cosmologically allowed range in $\sigma$ is covered by the two broken black curves obtained by varying the inflationary energy scale in the range $\rho_{\rm inf}^{1/4} = (1.120 - 1.216) \times 10^{15}$ GeV) with $N_c=4\pi$ fixed. This gives the range $\Lambda L_P^2=(2.657 - 3.056) \times 10^{-122}$, shown by the horizontal broken (blue) lines. Note that, because our results only depend on the combination $\rho_{\rm inf} L_P ^4 \exp(- 24 \pi^2 \mu)$, the same set of curves can also be incorporated in a Planck scale inflationary model ($H_{\rm inf} = L_P^{-1}$) with $\mu$ in the range 1.147-1.148. 
}
\label{fig:lambdasigmalowerr}
\end{figure}

We illustrate the above result graphically in Fig. \ref{fig:lambdasigmalowerr}. 
For the range of $\sigma $ allowed by cosmological observations, we get an acceptable range of values for $\Lambda L_P^2$ when $\rho^{1/4}_{\rm inf} =  (1.120 - 1.216)\times 10^{15}$ GeV). This shows, gratifyingly, that for well accepted models of inflation, and for the acceptable range of cosmological parameters, our postulate $N_c = 4\pi$ leads us to the correct value for the cosmological constant. (As we said before, this result  holds more generally with $\rho_{\rm inf}$ interpreted as $\rho_{\rm reheat}$ viz. the energy density at the end of reheating.)

\textit{We emphasize that there is no a priori reason to expect this result to hold.} Once we postulate $N_c=4\pi$ and use the observationally determined values for $\sigma$, we could have found that the inflation scale comes out to be an unreasonable number like, say,  $E_{\rm inf}=10^{10}$ GeV or $E_{\rm inf}=10^{21}$ GeV. Then we could not have reached sensible conclusions. \textit{The fact that everything works out consistently suggests that some deeper principle is in operation.}

\begin{figure}
 \begin{center}
  \includegraphics[scale=0.85]{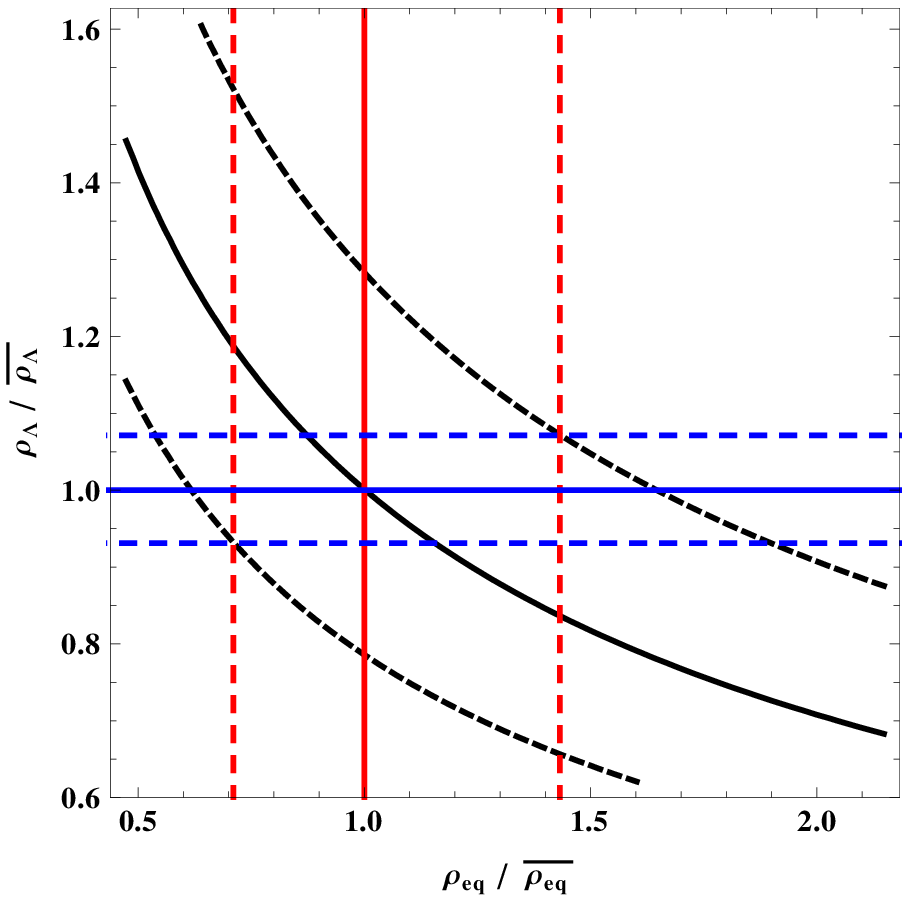}
 \end{center}
\caption{This figure shows the determination of $\rho_{\Lambda}$ from the range of $\rho_{\rm eq}$ permitted by cosmological observations,  with the postulate that $N_c=4\pi$. 
The values in the $y$-axis are normalized with respect to the observed value of the \cc, $\bar{\rho}_{\Lambda} L_P^4 =  1.135 \times 10^{-123}$ for simplicity. Similarly, the $x$-axis values of $\rho_{\rm eq} L_P^4$ are normalized with respect to the observed value $\bar{\rho}_{\rm eq} L_P^4 = 2.409 \times 10^{-113}$ indicated by cosmological observations.
The thick black curve is obtained from \eq{ll6} if we take the energy scale of inflation to be $\rho_{\rm inf}^{1/4} = 1.166 \times 10^{15}$ GeV in natural units. The cosmologically allowed range in $\rho_{\rm eq}$ is bracketed by the two broken black curves obtained by varying the inflationary energy scale in the range $\rho_{\rm inf}^{1/4} = (1.120-1.216) \times 10^{15}$ GeV.
This gives the range $\rho_{\Lambda} L_P^4  = (1.057 - 1.216) \times 10^{-123}$, which is consistent with observational results.}
\label{fig:rholvsrhoeq}
\end{figure}

In \eq{holy2}, we use the set $(\rho_{\rm inf}L_P^4,\sigma,H_\Lambda)$ as the independent epoch-invariant parameters determining the evolution of the universe and our postulate $N_c=4\pi$ determines one of them (viz. $H_\Lambda$), in terms of the other two. This set of parameters are directly amenable to putting observational constraints. However, it is also possible to express the same idea in terms of the set of epoch-invariant parameters $(\rho_{\rm inf} L_P^4, \rho_{\rm eq} L_P^4, \rho_\Lambda L_P^4)$ which is more in tune with fundamental physics.
In this case, our postulate 
$N_c=4\pi$ determines $\rho_\Lambda L_P^4$ in terms of the other two energy densities $\rho_{\rm eq} L_P^4$ and $\rho_{\rm inf} L_P^4 $. The exact expression is slightly complicated because it involves the solution to a quartic equation but for the cosmologically relevant range, the result is very well approximated by the relation in \eq{ll4} with $N_c=4\pi$. That is, we can express $\rho_\Lambda$ as a function of $\rho_{\rm eq}$ and $\rho_{\rm inf}$ as:
\begin{equation}
 \rho_\Lambda  \approx \frac{4}{27} \frac{\rho_{\rm inf}^{3/2}}{\rho_{\rm eq}^{1/2}} \exp (- 36\pi^2)
\label{ll6}
 \end{equation} 
The above result is plotted in Fig. \ref{fig:rholvsrhoeq}. 

When viable particle physics models for (i) GUTs scale inflation,  (ii) the dark matter candidate and (iii) baryogenesis are available, thereby determining  the ratios $n_m/n_\gamma,n_{\rm B}/n_\gamma, m_{\rm B}/M_P, m_{\rm DM}/M_P$ (where $M_P$ is the Planck mass), we can determine the value of $\rho_{\rm eq}$ [see \eq{heprhoeq}] and $\rho_{\rm inf}$ from first principles \cite{mazumdar}. Then, \eq{ll6} allows us to determine the numerical value of the \cc.
 In \eq{ll6}, the separation of effects due to the matter sector of the theory is evident. (Incidentally, the  $e$-folding
factor during the inflationary era in $PX$, when the cosmologically relevant modes exit the horizon, is given by the geometrical factor $6\pi^2\approx 60$ in this case.)

Before we conclude this section, we briefly comment on a scenario involving Planck scale inflation. To do this, we rewrite \eq{holy3} as:
\begin{equation}
 \Lambda L_P^2 = \frac{8 \pi \rho_{\rm inf} L_P^4 C(\sigma)}{3} \exp[-24\pi^2\mu];\quad \mu=\frac{N_c}{4\pi}
\label{holy4}
\end{equation} 
and note that for a given value of $\sigma$,
 $\Lambda L_P^2$ in \eq{holy4} depends only on the combination $\rho_{\rm inf} L_P^4 \exp(-24\pi^2\mu)$. So, we get the same value of $\Lambda L_P^2$ (for a given $\sigma$) when this factor has a given value. Among all possible choices, $\mu=1$, and $8 \pi \rho_{\rm inf} L_P^4/3 = 1$ (implying $H_{\rm inf} L_P = 1$), being natural, deserve special attention and we just completed discussion of the $\mu=1$ case. We next consider the $H_{\rm inf} L_P = 1$ case
which  corresponds to  Planck scale inflationary models. When $\sigma $ varies in the range allowed by observations, we find that $\mu$ varies between 1.147 and 1.148. Because $\Lambda L_P^2$ in \eq{holy4} involves $\mu$ in the exponential, one might be skeptical about the narrowness of this range of variation. However, the Planck scale inflation modeled with $H_{\rm inf} L_P = 1$ has  three conceptually attractive features. 
 
 First, the result is (trivially) independent of $H_{\rm inf}$ (since it is set to $L_P^{-1}$) and gives a direct relation between the two length scales $\Lambda$ and $L_P^2$ which occur in the gravitational physics of the universe. The  dependence of the result  on $\sigma$ is very weak and could be thought of as a higher order correction (like, for example, the   correction   beyond Bohr's model that leads to the fine structure of the spectral lines). 

Second, in such a model, we think of the  evolution in the intermediate phase (about which most cosmological investigations are concerned with!) as a mere transient connecting two de Sitter phases, both of which are semi-eternal. The fact that the de Sitter universe is time-translationally invariant makes it the natural candidate to describe the geometry of the universe dominated by a single length scale  --- which is $L_P$ in the initial phase and $\Lambda^{-1/2}$ in the final phase. The quantum instability of the initial de Sitter phase  can lead to cosmogenesis (see e.g., \cite{cosmogenesis}) and the transient radiation/matter dominated phase, which  eventually, gives way to the  late-time acceleration phase.  

Third, a theoretical  argument for $N_c \approx 4\pi$, which we will provide in Sec.\ref{sec:holoequi}, is quite natural with Planck scale inflation. It is obvious that the transition at $X$ is entrenched in Planck scale physics in such a model, which can easily account for deviations of $\mu$ from unity. We believe this model deserves study in the context of candidate models of quantum gravity.
 (As an aside, we mention the following: It is sometimes claimed  in the literature that Planck scale inflation is ruled out because it produces too much of gravitational wave perturbations. What can be actually proved is that, \textit{continuum} quantum field theory of spin-2 perturbations during inflation will lead to unacceptably large gravitational wave background, if the inflation scale is close to the Planck scale. But  one cannot really \cite{tpdetcc} trust \textit{continuum field theory} of the spin-2 field at Planck scales, based on which this result is obtained. There are  suggestions \cite{TPseshTP} that this problem vanishes if corrections to  propagators arising from a cut-off at the Planck scale are included.)

In fact, the simplest model for the universe, from this perspective, is the one with Planck scale inflation and just a radiation dominated phase in the intermediate stage. (That is, we ignore the matter dominated phase which exists for just 4 decades out of some 28 decades of expansion of the universe from the end of inflation till the beginning of late time acceleration.)
In other words, we approximate the intermediate phase of the  universe as purely radiation dominated (so that $\sigma\to\infty$) and assume Planck scale inflation (so that $H_{\rm inf} L_P=1$) thereby eliminating \textit{all} the free parameters from the theory. The above procedure now \cite{tpdetcc} gives  $\Lambda=\Lambda(N_c,\sigma^{-1}=0,H_{\rm inf} L_P =1)$ as a function of just $N_c$, as: 
\begin{equation}
 \Lambda L_P^2 = \frac{3}{4} \exp(-24\pi^2 \mu); \qquad \mu \equiv \frac{N_c}{4\pi} \, .
\label{holy1}
\end{equation}
Thus, $\Lambda L_P^2$ is expressible directly in terms of \cn\ and, in this case, there are \textit{no other adjustable parameters}.   
     \eq{holy1}  gives the observed value $\Lambda L_P^2 = 2.85\times 10^{-122}$ when $\mu =1.147$. (Incidentally, the same analysis works whenever there is only a  a single matter species; if $\rho\propto a^{-n}$, the exponential factor becomes $\exp[-12\pi^2\mu(n/(n-2))]$. The radiation dominated model $n=4$ gives the value of $\mu$ closest to unity.)
This shows that we are close to the correct value for $\Lambda L_P^2$ even in this simplest possible model. It is gratifying that adding GUTs scale inflation and the matter dominated phase
allows a completely consistent description.

\subsection{Discussion of the result}\label{sec:disc}

The evolution of the universe is treated in a rather fragmentary manner in the standard approach to cosmology, mainly because of historical reasons.  Planck scale physics is governed by the length scale $L_P$;  the inflationary scenario introduces $\rho_{\rm inf}$, while the  epoch of radiation and  matter sector domination introduces $\rho_{\rm eq}$; finally, to describe the   late-time acceleration, we need to introduce $\rho_\Lambda $. All these parameters are independently specified, with no relation with each other. Even assuming that GUTs scale physics will eventually determine $\rho_{\rm inf}$ and $\rho_{\rm eq}$ through \eq{heprhoeq}, we still need a link between the parameters, $L_P$ and $\rho_\Lambda$. In this context, the following aspects are worth emphasizing:
\begin{itemize}

\item
The \eq{ll6} directly links the inflationary energy scale $\rho_{\rm inf}$ and $\rho_{\rm eq}$ with the value of the \cc. This is a prediction of the model and, for it to work, the reheating scale after inflation scale is constrained to the range of about $\rho_{\rm reheat}^{1/4} = (1.12 - 1.22) \times 10^{15}$ GeV. If a given inflationary model reheats efficiently, this leads to the result $\rho_{\rm reheat}^{1/4}=\rho_{\rm inf}^{1/4} = (1.12 - 1.22) \times 10^{15}$ GeV. If the reheating is inefficient, then $\rho_{\rm reheat}^{1/4}$ will be lower that $\rho_{\rm inf}^{1/4}$ by about a factor 10-15. [Many inflationary models lead to an e-folding uncertainty of $\Delta N \approx 3$ or so, during which time the equation of state factor changes from $p/\rho\approx-1$
to $p/\rho\approx+(1/3)$. The dilution of the energy density scale will be  by a factor of about $ \exp (-\Delta N)$, which could be about 10-15.] Hence, if the model is right, it \textit{predicts} that the inflationary energy scale is about:
\begin{equation}
 \rho_{\rm inf}^{1/4} = (1-15) \times 10^{15} \ \text{GeV}
\end{equation}
Such a constraint is a strong prediction of the model, and hence the model is, in principle, falsifiable. In fact, if a specific inflationary model with all the details of reheating etc. are given, one can immediately determine whether it is consistent or not.

\item In the  paradigm introduced here, the postulate $N_c=4\pi$ acts as the connecting thread leading to a holistic approach to cosmic evolution, as is evident from \eq{ll6}. In a consistent quantum theory of gravity, we expect
 inflation (which determines $\rho_{\rm inf}$) and genesis of matter (which determines $\rho_{\rm eq}$ through \eq{heprhoeq} in terms of $n_{\rm DM}/n_\gamma,n_{\rm B}/n_\gamma$ etc) to be related to Planck scale physics such that our fundamental relation in \eq{ll6}  holds. Note that this is equivalent to the condition:
\begin{equation}
 \rho_{\rm eq} \rho_{\rm inf}^{-3} = \text{constant}
\end{equation} 
which is a relation between the inflationary scale and the matter/radiation created in the universe. This is a prediction of the model which is equivalent to the prediction of the inflationary scale when $\rho_{\rm eq}$ is taken as an input from observations.

\item In this approach, we solve the cosmological constant problem \textit{by actually determining} its numerical value in terms of other  parameters. As far as we know, such an approach to this problem has not been attempted before. 
This approach may be thought of as being similar in spirit to the Bohr model of the hydrogen atom, in which  the  postulate of $J=n\hbar$ was used to explain the observed energy levels of hydrogen. Here, our postulate $N_c=4\pi$, provides a connecting link between the three phases of evolution and explains the observed value of $\Lambda L_P^2$.
This postulate $N_c=4\pi$ is simpler and more powerful than many other ad-hoc assumptions made in the literature \cite{ccreviews} to solve the cosmological constant problem. 

\item \textit{More importantly,  this postulate is indeed correct!}  As we described in Sec.\ref{sec:cosmin},  the value of \cn\ 
can be determined directly from the observed value of $\Lambda$ and other cosmological parameters. Figs. \ref{fig:muwithcosmo},\ref{fig:muwithbeta} show that
$N_c$ is indeed very close to $4\pi$.

\item In the standard inflationary paradigm, the crossing of the Hubble radius by modes has \textit{no}   physical significance and is merely a simple way to describe the behaviour of the perturbation equation in two limits. For example, instead of the crossing condition $aH=k$,  one could have used $k/\pi,k/2\pi,....$ on the right hand side. The importance of \cn\ is probably related to the the cosmic parallelogram $PXQY$ (see Fig.\ref{fig:hubble}) which arises \textit{only} in a universe having three distinct phases. The epochs $P$ and $Q$, limiting the otherwise semi-eternal de Sitter phases, now  have a special significance \cite{bj,tpbj1,tpbj2}.  Modes which exit the Hubble radius before $a=a_P$ never re-enter. On the other hand, the epoch $a=a_Q$ denotes (approximately) the time when the CMB temperature falls below the de Sitter temperature \cite{bj,tpbj1,tpbj2}. The special role of $PXQY$ makes the value of  \cn\ (which is the same for $PX$, $XY$ or $YQ$) significant. 
It is, however, worth mentioning that in the alternative coordinate system used in Appendix \ref{appen:FRWrhop} to describe the  Friedman geometry the Hubble radius plays a crucial role. This will be of importance in our discussion in Sec.\ref{sec:egandcc}. 

\item As shown in Fig.\ref{fig:hubble}, these modes in $PXQY$ (with $k_-<k<k_+$) cross the Planck length during $a_-<a<a_+$, and it is likely that Planck scale physics imposes the condition $N(a_-,a_+)\approx 4\pi$ at this stage in the correct quantum cosmological model. As we shall see, it is possible to reformulate the field equations of gravity and the evolution equation for the universe in terms of the surface and bulk degrees of freedom. In this approach, one  attributes $(A/L_P^2)$ degrees of freedom with the area $A$. So, when the relevant proper radius of a Hubble sphere is $L_P$, one would associate $4\pi L_P^2/L_P^2=4\pi$ as the relevant parameter describing the degrees of freedom. It appears that this number is imprinted in the cosmic evolution and determines the value of \cn.
Clearly, this invites further work  to examine  the role of  \cn\  and its numerical value from quantum gravitational considerations. We will now describe several aspects of this idea. 

\end{itemize}

\section{The emergent gravity paradigm and an alternate perspective on cosmology}
\label{sec:egandcc}

We argued in Sec.\ref{sec:nscc} that in order to solve the \cc\ problem we need an alternative approach to gravity which satisfies conditions (1) and (2) listed in page \pageref{page:conditions}. This is successfully taken care of in the emergent gravity paradigm, in which the gravitational field equations can be obtained from a thermodynamic variational principle.
In this approach, we associate thermodynamic potentials with all null vector fields in a spacetime. The maximization of the relevant thermodynamic potential (like entropy, free energy, ...) associated with \textit{all} the null vectors simultaneously, then leads to \eq{tf}. Maximization involves varying the null vector fields rather than the metric, and therefore, it sidesteps  the third requirement in our list in page \pageref{page:conditions1}.
The metric is not varied to obtain the field equations.  In this approach,  the original variational principle, and not just the field equations, is invariant under the transformation in \eq{tsym}.
Such a variational principle can be obtained by using the functional \cite{grtd, aseemtp}:
\begin{equation}
 Q\equiv \int_{\lambda_1}^{\lambda_2} \frac{d\lambda \ d^2x}{16\pi}\, \sqrt{\sigma}\, [2\mathcal{S} + 16\pi T_{ab}\ell^a\ell^b ]
\end{equation}
where
\begin{equation}
 \mathcal{S}\equiv [\nabla_i \ell^j \nabla_j\ell^i -(\nabla_i\ell^i)^2 ]
\end{equation} 
can be shown to be
 the heat (enthalpy) density associated with the null surface containing the null congruence $\ell^a$. 
Since $T_{ab}\ell^a\ell^b$ can be interpreted as the heat (enthalpy) density $(\rho+p)=Ts=TS/V$ of matter, we can again interpret the integrand as the total heat density of the null surface.
Extremisation of the above functional for all null surfaces, then leads to \eq{tf} and satisfies all our criteria; and the \cc\ arises as an integration constant in this approach.

If these ideas are correct, then it must be possible to reformulate the field equations of gravity entirely in a thermodynamic language and do away with the standard geometrical description, based on, say, $G^a_b=\kappa T^a_b$. This is indeed possible and the final equations --- which, as we noted before, are identical to Einstein's equations with a \cc\ --- can be given a completely holographic interpretation in this approach, thereby giving us greater insight into the evolution of spacetime. We shall now describe how this comes about.

\subsection{Concept of Holographic Equipartition}

The degrees of freedom are the basic entities in physics. One possible way of introducing the  holographic principle is to relate  the number of degrees of freedom $N_{\rm bulk}$ residing in a bulk region $\mathcal{V}$ of space and the number of degrees of freedom $N_{\rm sur}$ on the boundary $\partial\mathcal{V}$ of this region. For a surface of area $A$, the surface degrees of freedom are counted as  $A/L_P^2$ where  $L_P^2$ acts as a fundamental area  scale. So, we shall define:
\begin{equation}
 N_{\rm sur}\equiv\frac{A}{L_P^2}=\int_{\partial \mathcal{V}} \frac{\sqrt{\sigma}\, d^2 x}{L_P^2}
\end{equation} 
where $\sigma$ is the determinant of the induced metric on $\partial\mathcal{V}$.
The non-trivial task is to come up with a suitable measure for the bulk degrees of freedom which must depend on the matter residing in the bulk. (This necessary dependence on the matter variables precludes counting the bulk degrees of freedom as $V/L_P^3$.) 
It is here that the idea of equipartition comes in. 

We begin by noting that one can associate a  local spacetime temperature $T_{\rm loc}$ with every point on the surface $\partial\mathcal{V}$, (which --- in general --- could vary on the surface) using the concept of local Rindler observers and local Rindler temperature, introduced as follows: Let $u^i$ be the four-velocity of fundamental observers with $x^\alpha$ = constant in a spacetime. Such observers will (in general) have an acceleration $a^i=u^j\nabla_ju^i$ with  respect to the freely falling observers at the same event. In the local inertial frame attached to the freely falling observer, one can think of the fundamental observers as equivalent to local Rindler observers  with this acceleration, and the corresponding \cite{daviesunruh} Davies-Unruh temperature $T_{\rm loc}= Na/2\pi$ where $a^2=a_ia^i$ and $N$ is the lapse function which takes care of the Tolman factor for the temperature. This defines a natural temperature at each event using the fundamental observers in a given coordinate system.

We can then assign an \textit{average} temperature $T_{\rm avg}$ to the surface $\partial\mathcal{V}$ by:
 \begin{equation}
 T_{\rm avg}\equiv\frac{1}{A}\int_{\partial \mathcal{V}} \sqrt{\sigma}\, d^2 x\ T_{\rm loc}
\label{tav}
\end{equation} 
 We can then think of $N_{\rm bulk}=|E|/(1/2)k_BT_{\rm avg}$ as the number of \textit{effective} bulk degrees of freedom where $E$ is the total energy in the bulk region $\mathcal{V}$ contributing to gravity. \textit{If}  the energy $E$ in the bulk region has reached equipartition with the surface temperature, \textit{then}, this is indeed the correct count of bulk degrees of freedom. So we define:
\begin{equation}
N_{\rm bulk}\equiv \frac{|E|}{(1/2)k_BT_{\rm avg}}= \pm\frac{1}{(1/2)k_BT_{\rm avg}}\int_\mathcal{V} \sqrt{h} d^3x\; \rho
\label{nbulkgen}
\end{equation} 
where $h$ is the determinant of induced metric on $\mathcal{V}$ and $\rho$ is the energy density of gravitating matter given by the Komar energy density, defined by the same fundamental observers:
\begin{equation}
\rho\equiv\rho_{\rm Komar} \equiv 2N [T_{ab} - (1/2)T g_{ab} ] u^au^b
\end{equation} 
We anticipate the possibility that $E$ could be negative, in which case, we will use the minus sign in the definition in \eq{nbulkgen} to keep $N_{\rm bulk}$ positive.
Given the expressions for $T_{\rm loc}$ and $\rho$, we can use \eq{tav} and \eq{nbulkgen} to determine $N_{\rm bulk}$ in the region $\mathcal{V}$.

Having defined the bulk and surface degrees of freedom in any spacetime, we next introduce the concept of holographic equipartition, which is the demand that 
\begin{equation}
 N_{\rm sur} = N_{\rm bulk} \qquad ({\rm Holographic \ equipartition})
 \label{key1}
\end{equation} 
Substituting the expressions for $N_{\rm sur}$ and $N_{\rm bulk}$, this reduces to the demand:
\begin{equation}
 \int_{\partial \mathcal{V}} \frac{\sqrt{\sigma}\, d^2 x}{L_P^2} \left( \frac{1}{2}k_BT_{\rm loc} \right)= \int_\mathcal{V} \sqrt{h} \ d^3x \ (\rho_{\rm Komar})
\label{holeq}
 \end{equation} 
An \textit{arbitrary} spacetime obeying the gravitational field equations will not, of course, be in holographic equipartition. But if the idea has to have some general validity, we would expect such an ``equilibrium'' condition to hold in any \textit{static} spacetime. This is indeed true and we shall first describe this  result briefly \cite{tpreviews,tpsurface}.

  Consider any static spacetime with a timelike Killing vector field $\xi^a = \delta^a_0$ and fundamental observers who are at rest in the spacetime with four-velocity $u^a = \xi^a/N$. Let the acceleration of these observers be $a^i = u^j \nabla_j u^i = (0, a^\mu)=(0,\mathbf{a})$. 
 At any given event $\mathcal{P}$, the static observers will experience  an acceleration $\mathbf{a}$ with respect to the freely falling observers at the same event, with  the corresponding Davies-Unruh temperature being $T_{\rm loc}= N|\mathbf{a}|/2\pi$. 
 Consider a closed region $\mathcal{V}$ in the 3-dimensional space \textit{bounded by an equipotential surface} $\partial \mathcal{V}$ such that the spatial normal to the surface $n_\mu=a_\mu/|\mathbf{a}|$ is in the direction of the acceleration. One can now  prove the holographic equipartition for any such bounded region in the form of \eq{holeq}.
Thus, \textit{gravitational field equations in any static spacetime imply holographic equipartition} \cite{tpsurface}. 

When the isothermal surfaces defined by $T_{\rm loc} = $ constant coincide with equipotential surfaces defined by the normal being in the direction of the acceleration, we can take the factor $(1/2)k_BT$ out of the integral in the left hand side and
the average temperature will be the same as the temperature of the isothermal surface. For example, this happens in all spherically symmetric spacetimes
in which the isothermal surfaces coincide with equipotential surfaces.

In fact, this result has an elegant generalization to an arbitrary, dynamic spacetime. We can show  \cite{grtd} that  the rate of change of gravitational momentum is related to the difference between the number of bulk  and boundary  degrees of freedom showing that the evolution of spacetime geometry is driven by the lack of holographic equipartition.
The dynamics of any spacetime can be described by the equation:
  \begin{equation}
 \int_\mathcal{V}\frac{d^3x}{8\pi}h_{ab}\pounds_\xi p^{ab}  = \frac{1}{2} k_B T_{\rm avg} ( N_{\rm bulk} - N_{\rm sur})
\label{start}
\end{equation} 
In the above equation, $h_{ab}$ is the induced metric on the $t=$ constant surface, $p^{ab}$ is its conjugate momentum and $\xi^a=Nu^a$ is the proper-time evolution vector corresponding to observers moving with four-velocity $u_a  = - N \nabla_a t$ which is the normal to the $t=$ constant surface. The terms $N_{\rm sur}$ and $N_{\rm bulk}$ are the degrees of freedom in the boundary and the bulk of a 3-dimensional region $\mathcal{V}$, respectively, and $T_{\rm avg}$ is the average Davies-Unruh temperature \cite{daviesunruh}  of the boundary, as before. The left hand side gives the time rate of change of the gravitational momentum, which, in turn, is driven by the departure from holographic equipartition, indicated by a non-zero value for $(N_{\rm bulk} - N_{\rm sur})$. The evolution ceases when $N_{\rm sur} = N_{\rm bulk}$ and, as we said before, all static geometries obey the above condition of holographic equipartition.  The validity of \eq{start} for all observers (i.e., for all foliations) assures the validity of Einstein's equations; in other words, \eq{start} carries  the same physical content as the gravitational field equations. In short, it is \textit{holographic equipartition that dictates the evolution of spacetime geometry}.

We shall now see how this prescription allows us to view cosmic expansion in a different perspective.

\subsection{The need for an alternative perspective to cosmology}\label{sec:coscon}

The standard approach to cosmology treats the evolution of the universe using the field equations of gravity, with our specific universe being selected out of all possible cosmological solutions by using observational inputs. To begin with, one assumes maximal symmetry so that the metric is described by a single function of time, $a(t)$, and a parameter $k=0,\pm 1$ which describes the spatial curvature. Next, one assumes a rather strange composition for the current universe (made of baryons, dark matter, radiation and dark energy) with the relative proportions being decided by certain numbers. The standard scenario for structure formation (and a few other issues) then also require an inflationary phase in the early universe with yet another form of source, possibly a scalar field.
These ad-hoc inputs, along with the field equations, then allow us to determine the evolution of our specific universe. 

The numerical coincidence described in Sec.\ref{sec:cosmin}, i.e. the fact that $N_c=4\pi$, \textit{does not 
 fit naturally} with this conventional paradigm. The fact that one can connect the three phases of the universe using $N_c$ and that one can determine the \cc\ using the postulate $N_c=4\pi$ calls for a deeper scrutiny of the conventional paradigm (unless we want to assume that it is merely a strange numerical coincidence, a point of view we find difficult to accept). The situation about the \cc\ is aggravated by the following two facts, described earlier.

\begin{itemize}
 \item As we saw in Sec.\ref{sec:nscc}, it is \textit{not} possible to solve the \cc\ problem unless (i) we can obtain the gravitational field equations in the form of \eq{tf} and (ii) interpret the \cc\ as an integration constant. This suggests that, to study cosmology, it is probably better to use an alternative perspective of gravity --- like the emergent paradigm --- which leads to \eq{tf}, than stick with the conventional paradigm.
 
 \item Even after we accept such a point of view, it is very difficult to make sense of the result $N_c=4\pi$ in any conventional approach. The definition of $N_c$ uses the Hubble radius in a significant and novel manner which has no analogue in conventional cosmology. We need to explore a broader context in which both the Hubble radius as well as the result $N_c=4\pi$ find a natural embedding. 
 
\end{itemize}

These aspects indicate that we should look for a suitable alternative postulate --- in place of the gravitational field equations --- which will allow us to determine cosmic evolution. Further, such a 
description should have the following features built into it:
(a) The Hubble radius should play a central role. The evolution of the universe is then better viewed as the dynamics of the Hubble volume. 
(b) The degrees of freedom, especially the notion of surface degrees of freedom on the Hubble sphere should play a key role in 
determining the dynamics, if we need to understand the connection with $N_c$. 
We will now describe an alternative paradigm which achieves these.

\subsection{Expansion as a quest for Holographic Equipartition}\label{sec:holoequi}

Let us get back to cosmology and see how these ideas help us to obtain an alternative description of cosmic expansion \cite{tpbj2}.
Consider first a de-Sitter spacetime described by a Hubble constant $H$. Such a spacetime has \cite{gh} a natural de-Sitter temperature $T = H/2\pi$. We now define the notion of surface and bulk degrees of freedom in a spherical region of radius equal to $H^{-1}$. For the surface degrees of freedom, we have
\begin{equation}
 N_{\rm sur} \equiv \frac{4\pi H^{-2}}{L_P^2}
\end{equation} 
which counts the number of area bits of size $L_P^2$ located on the surface.  As regards bulk degrees of freedom, we now have:
\begin{equation}
 N_{\rm bulk}=\frac{|E|}{(1/2) k_BT} = - \frac{2(\rho +3p)V}{k_BT} 
\label{Nbulk}
\end{equation} 
where $|E|$ is the magnitude of the  Komar energy $|(\rho +3p)| V$ contained inside the Hubble volume $V=(4\pi/3H^3)$. (We have used $|E|$ because $E$ is negative for the de Sitter spacetime and we want to keep $N_{\rm bulk} $ positive.)  Holographic equipartition is the demand in \eq{key1}, viz. that $N_{\rm sur} = N_{\rm bulk} $.
If we substitute $p=-\rho$, then \eq{key1} reduces to the standard result $H^2= 8\pi L_P^2 \rho/3$, for the de Sitter universe, showing that the de Sitter spacetime obeys the concept of holographic equipartition. (Here, we have used the \textit{proper} volume of the Hubble sphere $V = 4\pi/3H^3$ and the proper Komar energy density $(\rho +3p)$. One could have equally well used the  corresponding \textit{comoving} expressions which will differ by $a^3$ factors in both.) Thus, the demand of holographic equipartition leads to the same result as the gravitational field equations in this simple case. This should be obvious, in any case, because the de Sitter spacetime allows an alternative, spherically symmetric, static coordinate chart and we have already stated that holographic equipartition holds in all such spacetimes.
 
The situation really gets interesting when we move away from de Sitter and consider a general Friedmann model. 
  The result in \eq{key1} suggests that one can identify the de Sitter spacetime as some kind of an equilibrium state in which holographic equipartition holds. 
Our universe, of course, is not exactly de Sitter but there is 
considerable evidence that it is asymptotically de Sitter. This would suggest that when $N_{\rm sur} \ne N_{\rm bulk}$, the difference between them will drive the universe towards holographic equipartition. If this idea is true, then we would expect the difference ($N_{\rm sur} - N_{\rm bulk}$) to be the driving term for the cosmic expansion.  The  simplest form of such a law will be 
\begin{equation}
 \Delta V = \Delta t (N_{\rm sur} - N_{\rm bulk})
\label{key2}
\end{equation} 
 where $V$ is the Hubble volume in Planck units and $t$ is the cosmic time in Planck units. More generally, one would have expected $(\Delta V/\Delta t)$ to be some function of $(N_{\rm sur} - N_{\rm bulk})$ which vanishes when the latter does.
The \eq{key2} could be thought of as a Taylor series expansion of this function truncated at the first order. 
An alternative description of cosmic expansion, \textit{which does not begin from the field equations of general relativity,} can be obtained by elevating this relation to the status of a postulate. We will now show that this relation  is equivalent to the standard Friedmann equation. 

Reintroducing the Planck scale and writing $(\Delta V/\Delta t ) = dV/dt$, \eq{key2} becomes
\begin{equation}
 \frac{dV}{dt} = L_P^2 (N_{\rm sur} - N_{\rm bulk})
\end{equation} 
Substituting $V= (4\pi/3H^3), \ N_{\rm sur} = (4\pi/L_P^2 H^2), \ T=H/2\pi$ and using $N_{\rm bulk}$ in \eq{Nbulk}, this reduces to the relation:
\begin{equation}
 \frac{\ddot a}{a}=- \frac{4\pi L_P^2}{3} (\rho + 3p)
\label{frw1}
\end{equation} 
which is the standard accelerating universe scenario if we use the energy conservation for matter in the form $d(\rho a^3) = -pda^3$ and the de Sitter boundary condition at late times. 

The definition of $N_{\rm bulk}$ given in \eq{Nbulk} assumes that  $(\rho + 3p) <0$, thereby making $N_{\rm bulk}>0$. For normal matter, the negative sign in \eq{Nbulk} should be absent. This is easily taken care of by using appropriate signs for the two cases and modifying the equation to the form: 
\begin{equation}
 \frac{dV}{dt} = L_P^2 (N_{\rm sur} -\epsilon N_{\rm bulk}); 
\label{key3}
\end{equation} 
with the definition
\begin{equation}
 N_{\rm bulk} = -\epsilon \frac{2(\rho +3p)V}{k_BT} 
\label{nep}
\end{equation} 
We take $\epsilon = +1$ if $(\rho + 3p )<0$ and $\epsilon = -1$ if $(\rho + 3p ) >0$. (We could have used the opposite sign convention for $\epsilon$ and omitted the minus sign in \eq{nep}; this convention maintains the form of \eq{key2} for the accelerating phase of the universe.)
Because only the combination $+\epsilon^2 (\rho+3p)\equiv (\rho+3p)$ occurs in $(dV/dt)$, the derivation of \eq{frw1} remains unaffected with $N_{\rm bulk}>0$ in all cases.

To understand \eq{key3} better, let us  
separate out the matter component, which causes deceleration, from the dark energy component which causes acceleration.
We assume that the universe has two components (matter and dark energy) with $(\rho + 3p)>0$ for matter and $(\rho + 3p)<0$ for  dark energy. Then, \eq{key3} can be written as
\begin{equation}
 \frac{dV}{dt} = L_P^2 (N_{\rm sur} + N_m - N_{\rm de})
\label{key4}
\end{equation} 
with each of  $N_{\rm sur}, N_m, N_{\rm de}$ being positive and $(N_m - N_{\rm de})=(2V/k_BT)(\rho + 3p)_{\rm tot}$.
Note that,  if we want $dV/dt \to 0$ asymptotically, then  a component with $(\rho + 3p ) < 0$ must exist in the universe. \textit{That is, a universe without a dark energy component  cannot reach holographic equipartition.}  (The equation of state for the dark energy component cannot also not be too  different from that of a cosmological constant if the universe should not go into a super exponential expansion.) 
In other words, asymptotic holographic equipartition \textit{requires} the existence of a cosmological constant. In the presence of the cosmological constant, the expansion will make $N_{\rm de}$ dominate over $N_m$ at late times, driving the  universe into an  accelerated expansion. Asymptotically, we will have $N_{\rm de}\to N_{\rm sur}$ 
and the rate of volume expansion $dV/dt$, will tend to zero in a de Sitter universe.

There are three more 
features worth emphasizing about \eq{key2}.
The first is the striking  simplicity of this equation;  it is remarkable that the standard Friedmann equation can be reinterpreted as an evolution towards holographic equipartition. In fact, this result, originally obtained \cite{tpbj2} for general relativity has now been generalized to wider class of theories \cite{llholo}. \textit{If the background paradigm is not correct, it is very difficult to understand why \eq{key2} holds in our universe. }

Second, this equation presents the  evolution towards holographic equipartition in a ``bit-by-bit'' increase in Planck units. When the cosmic time changes by one Planck unit, the increase in Hubble volume is given by $(N_{\rm sur} - \epsilon N_{\rm bulk})$. In Planck units, \eq{key2} has no adjustable parameters and suggests the possibility of  interpreting cosmic expansion in purely combinatorial terms.

Finally,  the form of \eq{key4} suggests that there must exist a deeper relationship between the matter degrees of freedom and the dark energy degrees of freedom. In any fundamental theory of quantum gravity, we expect matter and gravitational degrees of freedom to emerge together and hence such a relationship is indeed expected. This resonates well with the idea of \cn\ as a unifying concept in the three phases of the universe and its value being equal to $4\pi L_P^2/L_P^2=4\pi.$

\section{Conclusions}

The dynamics of the spatially flat universe can be characterized by three densities: $\rho_{\rm inf},\rho_{\rm eq}$ and $\rho_\Lambda$, in addition to one undetermined overall normalization constant for $a(t)$ which could be taken to be the value of the expansion factor at the epoch when $\rho_m=\rho_R$. In standard cosmology, there is no inter-relationship between the three densities
$\rho_{\rm inf},\rho_{\rm eq}$ and $\rho_\Lambda$.

It is generally believed that high energy physics models will eventually provide a first-principles estimation of both $\rho_{\rm inf}$ and $\rho_{\rm eq}$. But we have no clue so far as to which physical principle determines the value of  $\rho_{\Lambda}$. 
 We have shown that there exists a characteristic number $N_c$ for the universe (``\cn'') which counts the number of length scales which enter the Hubble radius during the radiation/matter dominated phase. We have argued  in Sec.\ref{sec:cosmin} that this conserved number can act as a unifying link between the early inflationary phase, the radiation/matter dominated phase and the late-time acceleration phase.
 For a generic universe described by an unrelated set of three densities $\rho_{\rm inf},\rho_{\rm eq}$ and $\rho_\Lambda$, the parameter \cn\ can take any arbitrary value.
\textit{One of our key results is the discovery that \cn\ for our universe is equal to $4\pi$ to a high degree of accuracy.} (See \eq{muvalue}.) We strongly believe that this is unlikely to be an accident and demands an explanation.

If this result, $N_c=4\pi$ is raised to a status of a postulate, it can be used to determine the numerical value of $\rho_\Lambda$ in terms of the other two densities $\rho_{\rm eq}$ and $\rho_{\rm inf}$ (see \eq{ll6}) which --- as we said before --- are very likely to be determined by high energy physics. In other words, we now have a  paradigm in hand in which all the numbers characterizing the universe are determined from first principles. There is no a priori reason for such an idea to give results consistent with observations, which, in fact, happens only for a narrow range of inflationary energy scales. As mentioned in the first item in Sec.\ref{sec:disc}, this range is likely to be  constrained only by a factor of about 15 even if we take into account uncertainties due to inflationary reheating. In any case, given a detailed model of inflation, the procedure for calculating \cn\ is well defined and one could check for the consistency of the idea when inflationary models are on firmer footing. 

Such a procedure to determine the numerical value of $\rho_\Lambda$ has deeper implication for the structure of gravitational theories.  In conventional models, the value of $\rho_\Lambda$ can change when an arbitrary constant is added to the matter Lagrangian. In such a case, any physical principle to determine the numerical value of $\rho_\Lambda$ becomes dubious, since the value of the \cc\ that acts as the source for  gravity can be changed by adding a constant to the matter Lagrangian. As we have pointed out in Sec.\ref{sec:nscc},
this \textit{is a generic problem} in a large class of attempts to ``solve'' the \cc\ problem and the \cc\ problem can be really solved \textit{only if} we have two separate key ingredients in our model: 

(a) The gravitational \textit{field equations} must be invariant under the addition of a constant to the matter Lagrangian (which results in the modification of the zero level of energy, as in \eq{tsym}). 

(b) At the same time, the \textit{solutions} to the gravitational field equations must allow for the inclusion of a \cc, and  
we must provide a new principle to determine its numerical value. 

It was known for a long time \cite{aseemtp,tpap} that the emergent gravity paradigm, leading to the field equations of the form in \eq{tf}, takes care of the ingredient (a) above. What was lacking was a new physical principle which could be used to determine the value of the \cc, which arises as an integration constant to the solution of \eq{tf}. Here, we have provided this second ingredient (b) in the form of our postulate $N_c = 4\pi$. 

As mentioned in Sec.\ref{sec:coscon}, it is difficult to reconcile a result like $N_c=4\pi$ within the conventional cosmological paradigm which treats  the dynamics of the universe simply as a solution to the gravitational field equations. In fact, the conventional approach does not provide satisfactory answers to several other conceptual questions one could raise about our universe. For example,  one cannot even answer a simple question as to ``why does the universe expand'' within the context of standard classical cosmology \cite{TPwhy}. Since gravitational field equations are invariant under time reversal, one can write down the time-reversed solutions to the Friedmann equations describing a contracting universe.  All that we can  prove is that \textit{if} the universe is expanding today (which is taken as an observational input) \textit{then} it would have been expanding in the past --- though the initial singularity prevents us from meaningfully setting ``initial'' conditions to choose this solution. Further, in conventional cosmology, the universe seems to have evolved \textit{spontaneously} from a quantum mechanical state to a nearly classical state. It is not possible for a normal system to make a \textit{spontaneous} transition from a quantum to a classical state (in a rigorously defined sense, in terms of the Wigner function) if its evolution is governed by a bounded Hamiltonian  \cite{TPwhy}. Finally,  the Friedmann model breaks Lorentz invariance and chooses a preferred Lorentz frame (in which the CMB is isotropic), again because the solution to the field equations breaks the full symmetry of the field equations. All these features seem to suggest \cite{tpbj2} that we should describe cosmic expansion (and derive the Friedmann equations) from another physical principle, rather than treat them as arising as a solution to the gravitational field equations.

 We have argued in Sec.\ref{sec:coscon} that the concept of holographic equipartition provides such an alternate principle to describe cosmic expansion. This concept uses the Hubble radius fairly crucially and also identifies $4\pi$ as a primordial constant counting the number of degrees of freedom on a sphere of radius $L_P$. Both these ingredients go well with $N_c= 4\pi$ acting as a fundamental physical principle. Though the ideas presented in Sec.\ref{sec:coscon} of the review are still at a preliminary stage compared to the rest of the review, but they hold the promise of providing a novel and fruitful description of our cosmos.

\section*{Acknowledgements}

T.P's research  is partially supported by the J. C. Bose research grant of DST, India. 
H.P's research  is  supported by the  SPM research grant of CSIR, India.
We thank J. Bjorken, A. Lasenby,  D. Lynden-Bell, Don Page, J. Peacock, J. Sola for discussions and/or comments on a previous draft. These ideas were presented by one of us (TP) in the following two meetings: (a) Karl Schwarzschild Meeting, Frankfurt, July 2013 and (b) Cosmology and the Constants of Nature, Cambridge, March 2014. We thank several participants of these meetings for comments.

\begin{appendices}
\section{An alternate description of Friedmann geometry} \label{appen:FRWrhop}
 
 The description of the Friedmann geometry in \eq{s11.1}  uses the expansion factor $a(t)$, while Einstein's equations, \eq{s11.3}, determine $a(t)$ only up to a scaling factor. This has some interesting implications which are not adequately discussed in the literature. The purpose of this appendix is to highlight these features.
 
 There exists a different way of describing the geometry of the universe which is quite elegant and interesting. To introduce this description, we will proceed as follows: Consider a homogeneous, isotropic 3-dimensional space which, as we know, can be characterized completely by the spatial curvature. As in the main text, we shall assume that the spatial curvature is zero. This implies that the $t=$ constant sections of the universe are Euclidean. We can now introduce a radial coordinate $r$ in this Euclidean 3-space by the relation $r=(A/4\pi)^{1/2}$ where $A$ is the proper area of the $t=$ constant, $r=$ constant surface. It then follows that the angular part of the metric has the form $r^2[d\theta^2+\sin^2\theta d\phi^2]=r^2d\Omega^2$. Next, we introduce the energy density $\rho$ and $p$ as \textit{measured by the fundamental geodesic observers} (who see the 3-space as maximally symmetric) through the relations $\rho=T_{ab}u^au^b,p=(1/3)T_{ab}h^{ab}$ where $h_{ab}=g_{ab}+u_au_b$ is the projection tensor orthogonal to the four-velocity $u^a$ of the geodesic observers and $T_{ab}$ is the source energy-momentum tensor. Normally, to obtain the metric tensor, we now need to solve the Einstein's equations. However, \textit{this is unnecessary in the Friedmann universe} and the metric can be written out just in terms of $\rho$ and $p$, in the form:
 \begin{equation}
 ds^2 = - \frac{1}{24\pi G} \frac{d\rho^2}{\rho (\rho +p)^2} + \left[ dr + \frac{r}{3(\rho+p)} d\rho \right]^2 + r^2 d\Omega^2
  \label{prhometric}
 \end{equation}
 In the above metric, we are using the energy density in the universe $\rho$ itself as a locally defined time coordinate! The equation of state for matter gives $p$ as a function of $\rho$, and thus the line element in \eq{prhometric} is completely determined in terms of the `time coordinate' $\rho$. The fact that the Friedmann geometry can be described directly in terms of source variables without solving any differential equation is a nice feature which does not seem to have been noticed before in the literature. If we calculate the Einstein tensor for the metric in \eq{prhometric}, we will find that it satisfies Einstein's equations identically, with a source having energy density $\rho$ and isotropic pressure $p$ as measured by fundamental geodesic observers.
 
 It is easy to connect this up with the more conventional description. To do this, we introduce a new time coordinate by the definition:
 \begin{equation}
 t = - \left( \frac{1}{24\pi G} \right)^{1/2} \int \frac{d\rho}{(\rho+p)\sqrt{\rho}} 
  \label{prhotime}
 \end{equation}
 Using the geodesic equation for the metric in \eq{prhometric}, one can explicitly verify that $t$ is the proper time of geodesic clocks in the spacetime. Inverting this relation, we can express $\rho$ as a function of $t$, and using the equation of state, also obtain $p$ as another function of $t$. At this stage, it is also convenient to introduce the function $H(t)$ by the definition:
 \begin{equation}
 - \frac{1}{3} \frac{d\rho}{dt} \frac{1}{(\rho+p)} \equiv H(t) = \left( \frac{8\pi G \rho}{3}\right)^{1/2}
  \label{prhoH}
 \end{equation}
 where we have used \eq{prhotime} to obtain the second equality. The metric now takes a simpler form in terms of $t$ and $H(t)$ and is given by
 \begin{equation}
  ds^2 = - \left( 1 - H^2 r^2 \right) dt^2  - 2r H dr dt + \left( dr^2 + r^2 d\Omega^2\right)
  \label{H}
 \end{equation}
 In this form, the lapse function of the metric is unity, the shift function (in Cartesian coordinates) is $N^\alpha=-H(t)x^\alpha$ and the three-metric is trivial ($h_{\alpha\beta}=\delta_{\alpha\beta}$). 

There is wide-spread folklore (which is incorrect in the strict sense) that the dynamics of spacetime is encoded in the three-metric and the lapse and shift functions are purely kinematic in structure. We see that the dynamics of our universe can be \textit{entirely encoded} by the shift function with the lapse and the 3-metric being trivially unity.

 We can, of course, transform the metric in \eq{H} to the usual form in \eq{s11.1} by introducing a new function $a(t)$ and a coordinate $x$ by the relations:
 \begin{equation}
  H\equiv \frac{\dot a}{a}; \qquad x \equiv \frac{r}{a}
  \label{H1}
 \end{equation}
 This equation clearly shows that for a given $H(t)$ determined by the source, the corresponding $a(t)$ is not unique and is arbitrary with respect to a scaling by a constant. It is important to realize that such a scaling freedom does not exist if we use the coordinates in \eq{H} or \eq{prhometric} to describe the Friedmann geometry. If we rescale $a$, then the second equation in \eq{H1} tells us that $x$ is \textit{automatically} rescaled leaving $r$ fixed. We stress that $r$ has a direct geometrical meaning $(A/4\pi)^{1/2}$ in terms of the area of the $t=$ constant, $r=$ constant surface. So, the real origin of the scaling freedom in $a(t)$ is from the fact that the geometrical description in \eq{H} or \eq{prhometric} cares only for $H(t)$ and that the $a(t)$ arises through the definition in the first equation in \eq{H1}. 
 
 These facts also allow us to look at the notion of `expansion' of the universe somewhat differently. The description of Friedmann geometry in the coordinates in \eq{H} or \eq{prhometric} does \textit{not} make the 3-geometry expand! The metric of the $t=$ constant surface in \eq{H} is just the flat Euclidean metric with a constant 3-volume and the proper area of the $t=$ constant, $r=$ constant surface is \textit{independent} of time. In contrast, if we use the standard coordinates in \eq{s11.1}, the volume of the $t=$ constant surface varies as $a(t)^3$ and the area of the $t=$ constant, $r=$ constant surface varies as $a(t)^2$. \textit{Clearly, the `expansion' of the universe depends on the coordinates used to describe Friedmann geometry.} What is more, $g_{00}$ vanishes at the Hubble radius when we use the coordinate system in \eq{H}, which attribute greater importance to the Hubble radius.
 
 The key difference between the two coordinates is the following. The world lines with $r,\theta,\phi$ held constant are \textit{not} geodesics while the world lines with $x,\theta,\phi$  held constant are indeed geodesics. So, if we use \eq{H} as our metric, the geodesics describing fundamental observers are given by
 \begin{equation}
  r \exp \left[-\int H(t)dt\right] = \text{constant}
 \end{equation}
Since we believe galaxies are in geodesic motion in our universe, it is useful to use coordinates in which each galaxy has a constant value of
$x,\theta,\phi$ rather than a constant value of $r,\theta,\phi$. This is merely a question of convenience and not fundamental. 
 The geometrical features of our universe (like, e.g. redshift) do not change under coordinate transformations, and we can indeed talk of all physical phenomena using the metric in \eq{H} without ever introducing the notion of an `expanding' universe. We hope to return to this issue in a future publication \cite{hptpcosmvel}.

 \section{Comment on spatial curvature} \label{appen:commentkeq0}

In this appendix, we briefly comment on our assumption of $k=0$ in our calculations. If this assumption is relaxed, then the current value of the expansion factor, $a_0$, is given by the relation $a_0=H_0^{-1}\epsilon^{-1/2}$ where $\epsilon=|(\Omega_{tot}-1)|$. If inflation lasts for a sufficiently large period, $\epsilon$ will be extremely tiny. Even if we interpret the density of  gravitational waves generated during inflation as contributing to deviation from the spatially flat Friedmann model, we expect $\epsilon\approx\mathcal{O}(10^{-10})$ or so. When $k\neq0$, 
the Friedmann equation \eq{a1} gets modified to the form:
\begin{equation}
\left(\frac{\dot a}{a}\right)^2 =H_0^2\left[ \Omega_\Lambda +  
\Omega_m\frac{a_0^3}{a^3} +\Omega_R\frac{a_0^4}{a^4}-k\epsilon\frac{a_0^2}{a^2}
\right]\label{frwcurv}
\end{equation}
When $\epsilon$ is sufficiently small, the last term is dynamically negligible because the \cc\ dominates over it sufficiently early. To see this, note that the matter and dark energy densities are equal at $a=a_{m-de}$ where
\begin{equation}
 \left(\frac{a_{m-de}}{a_0}\right)^3=\frac{\Omega_m}{\Omega_\Lambda}
\end{equation} 
while the dark energy term and the last term in \eq{frwcurv} are equal at $a=a_{c-de}$ with
\begin{equation}
 \left(\frac{a_{c-de}}{a_0}\right)^2=\frac{\epsilon}{\Omega_\Lambda}
\end{equation} 
Therefore, the ratio
\begin{equation}
 \frac{a_{c-de}}{a_{m-de}}=\frac{\sqrt{\epsilon}}{\Omega_\Lambda^{1/6}\Omega_m^{1/3}}\approx
\mathcal{O}(\sqrt{\epsilon})
\end{equation} 
is extremely small. In other words, the dark energy starts dominating over curvature much before it dominates over the matter density. So, for all dynamical purposes, the curvature term is irrelevant in all phases of expansion for realistic values of $\epsilon$ arising from inflation. Hence, we consider a spatially flat Friedmann
model throughout the main text.
 
\section {Calculation of \cn} \label{appen:cncalc}

This appendix contains some  of the calculations leading to the results in \eq{eqformu}, \eq{ll4}, \eq{holy2} and \eq{holy4} in the main text. 

To study the evolution of a spatially flat universe containing a cosmological constant, pressure-free matter (which includes both dark matter and baryons) and radiation (which includes photons as well as any other relativistic species), one usually starts with the Friedmann equation written in the form of \eq{a1}. The Friedmann equation can further be written in terms of variable $x$, and the epoch-independent parameters $H_{\Lambda}$ and $\sigma$, as \eq{a6}.

Using this parameterization, we can now express the right hand side of \eq{Ndef} in terms of [$H_\Lambda, \sigma, \rho_{\rm inf}$].
Our postulate $N_c\approx 4\pi$ now gives a relationship among these parameters, inverting which we can express 
 $H_\Lambda $ (and thus the cosmological constant) in terms of $\rho_{\rm inf}$ and $\sigma$, both of which, of course, can be specified independently of $H_\Lambda$.  This is the strategy we will adopt.

We begin by determining $x_2$ and $x_1$, corresponding to $a_2$ and $a_1$ in \eq{Ndef}. It is easy to see that the condition $d[aH(a)]/da =0$, which determines $x_2$, allows us to express $x_2$ as a function of $\sigma$, say, as $x_2= r(\sigma)$, where the function $r(\sigma)$ satisfies the quartic equation: 

\begin{equation}
 2 \sigma^4 r^4 = 2 + r ; \qquad x_2 = r(\sigma) \, .
 \label{a8new}
\end{equation} 
 On the other hand, $x_1$ is determined by matching $H(a)$ with the Hubble constant during inflation, $H_{\rm inf}$. Since this occurs in the radiation dominated phase, we only need to retain the $x^{-4}$ term in the right hand side of \eq{a6}. This gives 

\begin{equation}
 H_{\rm inf}^2 \cong \frac{H_\Lambda^2}{\sigma^4 x_1^4};\qquad x_1^4 = \frac{H_\Lambda^2}{\sigma^4 H_{\rm inf}^2}
\label{a9}
\end{equation} 
We next compute $H_1^2x_1^2$ and $H_2^2x_2^2$. We have, from \eq{a9}:

\begin{equation}
 H_1^2 x_1^2 = \frac{H_{\rm inf} H_\Lambda}{\sigma^2}
\label{a10}
\end{equation} 
 Using \eq{a6}, we can express $H_2^2x_2^2$ as: 

\begin{equation}
 H_2^2 x_2^2 = H_\Lambda^2\left[ r^2+ \frac{1}{\sigma^4}\left( \frac{1}{r} + \frac{1}{r^2}\right)\right]
=\frac{ H_\Lambda^2}{2\sigma^4 r^2}\, \left[ 4  + 3r\right]\, ,
 \label{a11} 
\end{equation} 
giving the ratio:

\begin{equation}
 \frac{H_2^2 x_2^2}{H_1^2 x_1^2} = \left( \frac{H_\Lambda}{H_{\rm inf}}\right) \frac{(4+3r)}{2\sigma^2 r^2}\, .
 \label{a12}
\end{equation} 
Substituting this in \eq{Ndef}, and writing $N_c=4\pi \mu$, we get our final result:

\begin{equation}
 e^{-12\pi^2 \mu} =  \frac{H_\Lambda}{H_{\rm inf}} \, \left(\frac{4+3r}{2\sigma^2 r^2} \right)
 \label{a13}
\end{equation} 
This is equivalent to the result quoted in the text:
\begin{equation}
 \Lambda L_P^2 = \frac{8 \pi \rho_{\rm inf} L_P^4 C(\sigma)}{3} \exp[-24\pi^2\mu];\qquad 
\quad C(\sigma) = 6 (r + 2)\, (3r+4)^{-2}
\label{holy3app}
\end{equation} 
where   $r$ satisfies the quartic equation $\sigma^4 r^4 =   (1/2) r + 1 $.
Equivalently, we can express the above result for $\mu$ as:
\begin{equation}
\mu=\frac{N_c}{4\pi}=\frac{1}{24 \pi^2} \ln \left(\frac{8 \pi C(\sigma) \rho_{\rm inf} L_P^4}{3 \Lambda L_P^2} \right)
\label{eqformuapp}
\end{equation} 
which is \eq{eqformu} in the main text.

As we stressed before, $\sigma$ is an epoch independent parameter defined through \eq{a7} and its value is, of course, independent of $H_\Lambda$, which we are trying to determine. Since $\sigma$ is epoch independent, we can determine its possible value and the range from current observations. The quartic equation requires numerical solution and the result is plotted in \fig{fig:lambdasigmalowerr}. 

The quartic equation, \eq{a8new} can be solved approximately in two limits, viz., $\sigma \gg 1$ and $\sigma \ll 1$. 
In the first case ($\sigma \gg 1$) the approximate solution to the quartic in \eq{a8new} is given by $r\approx \sigma^{-1}$. In this limit, to the same  order of accuracy, \eq{a13} becomes 
$e^{-12\pi^2\mu} = 2 H_\Lambda /H_{\rm inf}$ which can be rewritten as 

\begin{equation}
 \Lambda L_P^2 = \frac{3}{4} \, (H_{\rm inf} L_P)^{2} \, \exp(-24\pi^2 \mu)
 \label{a14}
\end{equation} 
It is easy to verify that this limit corresponds to a purely radiation dominated universe with $\oms \to 0$. As mentioned earlier, our expression for $\Lambda$ can \textit{only} depend on $\oms$ and $\ors$ through $\sigma$;  when $\sigma \gg 1$, this dependence drops out. 

In the other limit of $\sigma \ll 1$, the quartic equation, \eq{a8new}, has the approximate solution  $r\approx (2\sigma^4)^{-1/3}$, and \eq{a13} gives:

\begin{equation}
  e^{-12\pi^2 \mu} = \frac{3 H_\Lambda}{H_{\rm inf}} \left( \frac{1}{2\sigma}\right)^{2/3}
\label{a15}
\end{equation} 
Raising both sides of the above equation to the third power, we obtain:
\begin{equation}
  e^{-36\pi^2 \mu} = \frac{9}{4}  \frac{\Lambda L_P^2}{(H_{\rm inf} L_P)^3} \, \left( \frac{H_\Lambda L_P}{\sigma^2}\right) 
 \label{a16}
\end{equation} 
Substituting $\Lambda L_P^2= 8 \pi \rho_{\Lambda} L_P^4$, $ H_{\Lambda} L_P = (8 \pi \rho_{\Lambda} L_P^4/3)^{1/2}$, $ H_{\rm inf} L_P = (8 \pi \rho_{\rm inf} L_P^4/3)^{1/2}$, and $\sigma = (\rho_{\Lambda}/\rho_{\rm eq})^{1/4}$ into the above equation, we obtain the relation between the three densities  $\rho_{\Lambda}$, $\rho_{\rm eq}$ and $\rho_{\rm inf}$ as quoted in \eq{ll4}:
\begin{equation}
 \rho_\Lambda  = \frac{4}{27} \frac{\rho_{\rm inf}^{3/2}}{\rho_{\rm eq}^{1/2}} \exp (- 36\pi^2 \mu) \equiv \frac{4}{27} \frac{\rho_{\rm inf}^{3/2}}{\rho_{\rm eq}^{1/2}} \exp (- 9\pi N_c)
\label{ll4new}
 \end{equation}

High energy physics models will eventually determine the inflationary energy density $\rho_{\rm inf}$, and the density at matter-radiation equality, $\rho_{\rm eq}$. Given these two numbers, and our postulate that $\mu = 1$, (i.e., $N_c=4\pi$), we can determine $\rho_{\Lambda}$ from the above equation, and thereby, the cosmological constant.

\end{appendices}

 \end{document}